\documentclass[aps,prd,amsmath,amssymb,showkeys,showpacs,10pt]{revtex4}
\usepackage{amsfonts}
\usepackage{graphicx}
\usepackage{bm}
\def\journal#1#2#3#4{{#1} {\bf #2}, #3 (#4)}

\newcommand{\be}{\begin{equation}}
\newcommand{\ee}{\end{equation}}
\newcommand{\bea}{\begin{eqnarray}}
\newcommand{\eea}{\end{eqnarray}}
\newcommand{\hf}{\frac12}
\newcommand{\nn}{\nonumber\\}
\def\eq#1{(\ref{#1})}
\def\la{\langle}
\def\ra{\rangle}
\def\Tr{{\mathrm{Tr}}}
\def\tr{{\mathrm{tr}}}
\def\mr#1{{\mathrm{#1}}}
\def\v#1{{\bm{#1}}}
\def\ord#1{{\cal O}(#1)}
\def\br{\hskip-6pt/}
\def\bre{\hskip-4pt/}
\def\fd#1#2{\frac{\delta#1}{\delta#2}}
\def\fdd#1#2#3{\frac{\delta^2#1}{\delta#2\delta#3}}
\def\ab{\bar a}

\def\psib{\bar\psi}
\def\ha{\hat a}
\def\hj{\hat j}
\def\hsigma{\hat{\sigma}}			
\def\htG{\hat{\tG}}
\def\hA{\hat A}
\def\hD{\hat D}
\def\hG{\hat G}
\def\hJ{\hat J}

\def\jb{\bar j}
\def\psib{\bar\psi}
\def\psid{\psi^\dagger}
\def\tG{\tilde G}
\def\Ab{\bar A}
\def\Jb{\bar J}

\newcommand{\TERMA}{\mathcal{A}}
\newcommand{\TERMB}{\mathcal{B}}

\begin{document}
\title{Sub-classical fields and polarization in electrodynamics}
\author{Mathieu Planat, Janos Polonyi}
\affiliation{Strasbourg University, High Energy Theory Group, CNRS-IPHC,
23 rue du Loess, BP28 67037 Strasbourg Cedex 2 France}
\date{\today}

\begin{abstract}
Expectation values of the electromagnetic field and the electric current
are introduced at space-time resolution which belongs to the quantum domain.
These allow us to approach some key features of classical electrodynamics 
from the underlying QED. One is the emergence of the radiation 
field in the retarded solution of the Maxwell equation, derived from an action
principle. Another question discussed is the systematic derivation of
the polarizability of a charge system. Furthermore, the decoherence and 
the consistency of the photon field are established by a perturbative calculation 
of the reduced density matrix for the electromagnetic field within the
Closed Time Path formalism. 
\end{abstract}
\maketitle

\section{Introduction}
Classical electrodynamics consists of a set of equations of motion for the 
electromagnetic (EM) field and the electric current, the Maxwell and the mechanical
equations, respectively. But the classical theory should be derived from QED rather than 
used as a starting point to quantize the EM field. When approached in this manner, these 
equations become approximate, govern the expectation values of the
EM field and the electric current, and are the subject of an infinite hierarchy of quantum corrections. 

The degrees of freedom common in quantum and classical physics are the expectation values
of local fields. But we have to keep in mind that such expectation values are not necessarily 
classical. The crucial quantity which controls the classicality of the field
expectation value is the number of degrees of freedom $N=n\xi^3$ within the 
internal distance scale of the field, the correlation length $\xi$ of a many-body system at
particle density $n$. The UV cutoff $\Lambda$, the maximal energy scale of QED, will be 
kept large but finite in this work: therefore, the internal scale of the field 
expectation value $\xi>1/\Lambda$ may fall into the quantum or classical domain, 
depending on initial conditions and external sources. For $N\sim1$ the local expectation value does 
not follow the laws of classical physics and will be called a subclassical field. 
Such fields are used for instance in the density functional method \cite{dreizler}.
These fields are useful to monitor the quantum-classical transition because they
bridge the gap between observables and expectation values, used on the microscopic 
and macroscopic domain, respectively. This can be seen simplest within the framework 
of the renormalization group where the classical action governing the dynamics 
of the subclassical field can be set as an initial condition for the integration 
of the renormalization group equations in the macroscopic domain.
This point of view motivated this work, the study of some necessary ingredients
of the quantum-classical transition by means of the dynamics of subclassical fields.
It includes the study of the dynamical origin of decoherence \cite{zehd,zurekd} 
and consistency \cite{grif}, the breakdown of the time reversal invariance 
\cite{zeh} and the way the expectation values are formed at the microscopic scale.

A reassuring feature of subclassical fields is that their dynamics is stable and 
well defined in the UV. In fact, the particles can not be localized with a precision better 
than their Compton wavelength therefore the subclassical fields are smeared out at shorter 
distances. This scale is independent of the underlying quantum field theory.
The subclassical field is classical in the sense that quantum fluctuations are 
averaged out. But it is still in the quantum domain due to its fine spatial resolution because 
the decoherence within this space region is not strong enough to restore classical 
probabilities, in particular, the additivity for exclusive events. 

The present work is the continuation of the program  outlined in Ref. \cite{pol} where the 
equations of motion for the expectation value of the EM field and the electric current were 
discussed in QED. We go beyond that work and render that approach more realistic by including 
a reservoir of charges at finite density and by calculating explicitly and by interpreting the vacuum 
polarization effects in the equation of motion for the EM field and in
decoherence/consistency. The loop expansion will be employed to include the vacuum polarization 
effects and we shall be satisfied by calculating the leading order, one-loop approximation
to the linearized variational equations of motion. The current terminology of the traditional
construction of the averaged fields in classical electrodynamics can be borrowed from the 
renormalization group method, it is a special blocking. In the present work we are 
satisfied by a simpler realization of this idea, the coarse graining generated
by the elimination of the charged degrees of 
freedom from the dynamics without lowering the space resolution. The decoherence and
consistency become simple perturbative effects in this setup, they are generated by the
collective modes of the many-body vacuum.
The dynamical breakdown of the time reversal invariance can be traced back to the collective 
modes as well, it is driven by the finiteness of the life-time of the quasiparticles.
Both the decoherence/consistency and dynamical breakdown of the time reversal invariance 
are actually generated by the same piece of the photon polarization tensor. This is how 
the dynamical breakdown of the time reversal invariance and decoherence appear coupled 
in the classical limit. Note that there is no need of a particular definition
of pointer states in this procedure, quasiparticles arising form our
coarse graining provide a natural,
dynamically generated robust basis in the Fock space for the study of the
reduced density matrix.

The breakdown of the time reversal invariance brings the boundary conditions in time 
into the foreground. The other goal of this work is the establishment of the
dynamical symmetry breaking and its clear separation from the explicit breakdown by
the boundary conditions in time. The not entirely trivial issue of the latter is the way the 
external sources rearrange the radiation field in the variation principle to comply with the given
boundary conditions. The simplest way to regulate the corresponding formal, 0/0 terms in the equations
of motion is the introduction of the usual $i\epsilon$ term in the quadratic part of the
action in classical field theory, as well as in its quantum counterpart.

The quantum corrections to the space-time dependence of the averages of local 
observables are usually calculated by means of the linear response formalism. 
We need a more powerful method, capable of dealing with the eventual nonperturbative issues
of the quantum-classical crossover regime to arrive at a reliable description
of quantum effects, turning classical electrodynamics into QED at short distances.
The Closed Time Path (CTP) method developed by J. Schwinger \cite{schw} 
provides us with a nonperturbative setting not only for the determination of 
the space-time dependence of local observables but for other key ingredients of 
the quantum-classical transition, to the reduced density matrix and the possible
dynamical breakdown of the time reversal invariance, too. 

The CTP method has been extended a number of times 
\cite{keldysh,leplae,zhou,su,dewitt,calzetta,arimitsu,umezawa,campos}.
Despite being the only way of obtaining expectation values as opposed to
transition amplitudes reproduced by conventional quantum field theory its 
inherent algebraic complications somehow limited its applications.
The generator functional for connected Green functions, our starting point given by 
Eq. \eq{waj} below is closely related to the influence functional introduced in 
Ref. \cite{feynman}. One may simple say that all that is presented below is the
application of the influence functional scheme to derive the effective action
and the equations of motion for the expectation value of local operators.
The solution of the latter provides us with a partial resummation of the perturbation
series with rich physical content. Furthermore, certain details of the influence
functional contain decoherence/consistency and irreversibility, the key conditions
for the classical limit. The reduced density matrix and dissipative forces have been 
studied in the framework of nonrelativistic quantum mechanics in a manner
reminiscent of the CTP formalism \cite{weiss}. The unusual feature of quantum transition 
amplitudes, the possibility of fixing the initial and the final states of the motion has 
already been addressed in a manner, equivalent to the CTP formalism in Ref. \cite{aharonov}. 
The Schwinger-Dyson equation, the removal of the UV divergences, the equation of motion 
for the EM field derived in the CTP formalism and the pair creation generated by a
homogeneous external electric field have already been thoroughly studied for a large 
number of charged fermions in Ref. \cite{cooper}. Our results are complementary 
since they retain the adiabatic, IR tail of the vacuum polarization only.
It is worth mentioning that the problem of establishing variational equations for
the expectation values \cite{jordan} can be resolved by a suitable parametrization
of the degrees of freedom \cite{pol}. The role of collective modes in the
quantum-classical transition can conveniently be studied within the CTP scheme \cite{coulomb},
as well.

We start in Section \ref{doubletime} with a short introduction of the CTP formalism,
including the systematic construction of the Heisenberg representation in terms of Green functions.
Instead of following the historical route, the motivation of this scheme by the
calculation of the expectation values rather than transition amplitudes, an alternative
argument is presented. The CTP formalism is motivated by the generalization of the 
classical action principle for initial condition problems which produce the retarded
solutions. Beside the CTP case we need the Open Time Path (OTP) 
version, as well, to determine the reduced density matrix. Some functionals, 
used subsequently are introduced here, too, namely those which generates the connected 
Green functions and their Legendre transform, the effective action. 

A rather general issue is addressed in Section \ref{fldres} where a nonrenormalization
theorem is shown for the Maxwell equations when all charges are included. 
One finds dressing only when charges are ignored. This result
is in agreement with the usual derivation of the dielectric constant and the
magnetic permeability in classical electrodynamics. 

The double-time formalism is introduced for QED in Section \ref{cgrefnct}, followed
by the calculation of the quadratic part of the generator functional for the
connected Green functions and the effective action for the electric current and the EM field,
as well as a short discussion of the ways the mass-shell modes of the photon and the zero modes of 
the free EM field action satisfy the desired boundary conditions in time by 
formal, 0/0-type terms in the equations of motion.The one-loop result for this functional 
is summarized in Section \ref{olplta}. 

The explicit calculation of the reduced density matrix for the EM field is
carried out in Section \ref{densmatr}. The real part of its logarithm controls 
the dielectric and the magnetic permeability functions and determines the quasiparticles,
the imaginary part is responsible for the finite life-time of the quasiparticles,
and the decoherence and the consistency of the EM field. Finally, Section \ref{concl}
summarizes our conclusions.

The Appendixes contain some technical details. A short summary of the CTP propagators 
in the vacuum is presented in Appendix \ref{ctppropa}, followed by the extension to
thermodynamical equilibrium in Appendix \ref{ctppropenv}.

\section{Time arrow and the double-time formalism}\label{doubletime}

The dynamical breakdown of the time reversal invariance is a provocative, not yet understood 
problem \cite{zeh}. Weak interactions set apart, the fundamental forces and dynamical
equations of Nature are invariant under a reversion of the direction of time. Nevertheless,
daily experiences show that external perturbations influence the systems at later
times only, and therefore the retarded solutions of the time reversal invariant equations of motion
should be used. We set up our quadrature to solve the differential equations accordingly,
by specifying the initial conditions. The resulting scheme obviously does not obey time reversal
invariance. Such an explicit breakdown of the symmetry is a trivial phenomenon, the 
real, dynamical issue is the simplicity of the initial conditions, and
the absence of the radiation field in electrodynamics or excited states, in general, 
in quantum mechanics.

Even this trivial issue poses an interesting question. How can such an initial 
condition problem be dealt with in physics? 
The Newton equation, being a second order differential equation, requires two input data
per degrees of freedom to fix a solution uniquely. For a differential equation
we can choose freely among parameters of the motion, either at the initial or at the 
final time. This allows us to treat problems where the initial or the final condition is
simple. A typical example is the use of the retarded or advanced Li\'enard-Wiechert 
potentials in classical electrodynamics. But the fundamental, most flexible layer of 
classical mechanics is the variational principle where boundary problems can be
defined only because the variation performed at the end point of the trajectory 
in an initial problem cancels the generalized momentum. Such a shortcoming of the 
variation method is not a serious problem in classical dynamics since we can
switch back to the canonical equations of motion anytime and set up the desired Cauchy problem. 
The situation is more involved in quantum mechanics, where the action principle remains important. 
In fact, after having set up a problem in the operator formalism, the actual 
calculation of the transition amplitudes or expectation values is usually done in 
the path integral formalism. Hence the goal of generalizing the variation principle 
to cope with retarded or advanced forces remains an important question.

How can we construct a variational method which automatically produces the retarded 
solution of the equations of motion? One should modify the procedure in such a manner
that it is enough to provide the coordinates and the velocities at $t=t_i$, the 
beginning of the time evolution. But what does one do with the variation of the 
trajectory at the final time $t=t_f$? A simple way to render the variation at the 
final time consistent with the classical trajectory is to cancel altogether its contribution 
to the equation of motion. This can be achieved by extending the dynamical problem
for twice as long as $0\le t_{CTP}\le2T=2(t_f-t_i)$. We follow the 
usual trajectory $x^+(t)$ for the time interval $T$ then we flip the time arrow 
and follow the time backward along another trajectory $x^-(t)$,
which returns to the initial conditions. The closed trajectory obtained in this manner is
\be
x^{CTP}(t_{CTP})=\begin{cases}x^+(t_i+t_{CTP})&0\le t_{CTP}\le T,\cr
x^-(2t_f-t_i-t_{CTP})&T\le t_{CTP}\le2T.\end{cases}
\ee
The variation problem for such a time reversed trajectory is well defined, and it yields
a trivial equation of motion 0=0 for $t=t_f$ and produces the retarded solution.
The quantum analogy of this scheme is well known, it is the CTP formalism
of Schwinger \cite{schw}. We shall need a slight generalization of this scheme, 
the OTP fromalism \cite{hig} to access the reduced density matrix 
of subsystems to cope with a genuine quantum effect, the decoherence.

The CTP formalism was originally developed as the standard perturbation expansion for expectation
values in the Heisenberg representation of Quantum Field Theory \cite{schw}. When Feynman's 
insight to QED, based on a simpler scheme of transition amplitudes became widely accepted,
then the perturbation expansion for the expectation values was gradually abandoned. This shift of
attention seemed to be further justified by the simplicity and generality of the path integral 
representation of the transition amplitudes. The interest in the CTP formalism was kept
in condensed matter physics, where expectation values corresponding to given, time-dependent
external perturbation are important. 

The expectation value
\be
\la\psi(t)|A|\psi(t)\ra=\la\psi_i|A_H(t)|\psi_i\ra,
\ee
obtained in the Heisenberg representation, where where $A_H(t)=e^{i(t-t_i)H}Ae^{-i(t-t_i)H}$ and 
$|\psi(t_i)\ra=|\psi_i\ra$ can be written as
\be\label{expv}
\la\psi(t)|A|\psi(t)\ra=\Tr[Ae^{-i(t-t_i)H}\rho_ie^{i(t-t_i)H}].
\ee
in terms of the initial density matrix $\rho_i=|\psi_i\ra\la\psi_i|$.
This expectation value can be related in a simple manner to the transition amplitude
\be
{\cal A}=\la\psi_i|e^{-i(t_f-t)H}Ae^{-i(t-t_i)H}|\psi_i\ra
\ee
only when the initial state $|\psi_i\ra$ is an eigenstate of the Hamiltonian. This condition is not
met in a realistic situation where the initial state contains complicated, collective 
excitations. Expectation values like \eq{expv} will be obtained in the CTP formalism, used 
in this work to derive the effective action whose Euler-Lagrange 
equation is satisfied by the expectation values of the EM field and the electric current.

\subsection{Generator functional for connected Green functions}
We recall first some general features of the CTP formalism and next we 
introduce this scheme for QED. The expectation value \eq{expv} motivates 
the introduction of the generator functional 
\be\label{eqwjBasique}
e^{iW[j^+,j^-]} = \Tr T[e^{-i\int_{t_i}^{t_f}dt\int d^3x [H(x)-j^+(x){\cal O}(x)]}]\rho_i
(T[e^{-i\int_{t_i}^{t_f}dt\int d^3x [H(x)+j^-(x){\cal O}(x)]}])^\dagger
\ee
where $j^\pm$ are two sources coupled to a local operator ${\cal O}$ and we use the
units $\hbar=c=1$. This generator
functional can easily be given in the path integral formalism. Indeed, the
time evolution operator to the left of $\rho_i$ has the standard path integral representation.
The same holds for the time evolution operator on the right except that
the replacement $j^+\to-j^-$ is carried out, as well. Note that the Hermitian
conjugation of the operator standing to the right of $\rho_i$ leads to anti-time ordering. 
Consequently, the number of degrees of freedom is doubled, 
$\phi^+$ and $\phi^-$ will be used to denote the CTP doublets, field 
trajectories for the left and right time evolution operator, respectively.
We therefore have pairs of trajectories in the full path integral
\be\label{eqPathIntBasique}
e^{iW[j^+,j^-]}=\int D[\phi_f]D[\phi^+_i]D[\phi^-_i]\Psi^*_0[\phi^-_i]\Psi_0[\phi^+_i]
\int_{\phi^\pm_{t_i}=\phi^\pm_i}^{\phi^\pm_{t_f}=\phi_f}D[\phi^+]D[\phi^-]
e^{iS[\phi^+]+iS_{CT}[\phi^+]-iS[\phi^-]-iS_{CT}[\phi^-]+ij^+{\cal O}^++ij^-{\cal O}^-}
\ee
where ${\cal O}^\pm$ is constructed by means of the field $\phi^\pm$, the 
integral measure $D[\phi^\pm]$ includes the field variables for $t_i<t<t_f$
and the space coordinate of $\phi$ is suppressed. 
The initial condition $\rho_i=|0\ra\la0|$
is used where $|0\ra$ stands for the noninteracting vacuum state with wave
functional $\Psi_0[\phi]$ and the trajectories $\phi^+$ and $\phi^-$ are 
joined at the final time due to the trace operation in Eq. \eq{eqwjBasique}.
The boundary conditions in time do not influence UV divergences therefore
the counterterms contained in $S_{CT}[\phi]$ are imposed separately
for each time axes. It is not difficult to check that the boundary conditions
in time, used in Eq. \eq{eqPathIntBasique} can be incorporated by the replacement
$S[\phi^+]-S[\phi^-]\to S[\hat\phi]=S[\phi^+]-S[\phi^-]+S_{BC}[\phi^+,\phi^-]$
where $S_{BC}$ is a quadratic expression of the fields taken at $t=t_i$ and $t=t_f$.
The scalar product $fg$ of space-time functions stands for space-time integration, 
$fg=\int d^4xf_xg_x=\int_xf_xg_x$, vector and CTP indices being summed, if 
necessary, eg. $jA=\sum_{\sigma=\pm}\int_xj^{\sigma\mu}_xA^\sigma_{\mu x}$ for the
minimal coupling. Note 
that the expectations values, calculated for unitary time evolution $j^+=-j^-$ are 
independent of the value of this final time $t_f$ as long as they are considered before $t_f$. 
The bonus of these apparent complications in the CTP formalism is a way to 
handle the dynamical breakdown of the time reversal invariance, decoherence 
and dissipative forces \cite{pol,coulomb}.

We consider in this work elementary fermions of electric charge $e_n$, 
$n$ being a flavor index, interacting with the EM field. For this kind of system, the quantities 
like \eq{expv} can be obtained by the use of the generator functional
\be\label{waj}
e^{iW[a^+,a^-,j^+,j^-]}=\Tr T[e^{-i\int_{t_i}^{t_f}dt\int d^3x [H(x)-\sum_n a^+_n(x)J_n(x)-j^+(x)A(x)]}]\rho_i
T^*[e^{i\int_{t_i}^{t_f}dt\int d^3x [H(x)+\sum_n a^-_n(x)J_n(x)+j^-(x)A(x)]}]
\ee
where $T^*$ denotes anti-time ordering, $j^\pm$ are two sources used to generate 
the EM field $A$ and $a^\pm$ are two vectorial sources coupled to the EM current 
$\psib\gamma^\mu\psi=J^\mu$. Eq. \eq{waj} can be written as 
\be\label{wpint}
e^{iW[\ha,\hj]}=\int D[\hA]\biggr(\prod_n D[\hat\psi_n]D[\hat{\psib}_n]\biggl)
e^{i\sum_n\sum_{\sigma\sigma'}\psib^\sigma_n[(G_{0n}^{-1})^{\sigma\sigma'}
+\delta^{\sigma\sigma'}(a\bre^\sigma_n-e_n\sigma A\bre^\sigma)]\psi^{\sigma'}_n
+\frac{i}{2}\hA\hD^{-1}_0\hA+i\hj\hA+iS_{CT}}
\ee
using the notation
\be
\hat\psi_n=\begin{pmatrix}\psi_n^+\cr\psi_n^-\end{pmatrix},~~~~~~
\hA=\begin{pmatrix}A^+\cr A^-\end{pmatrix},~~~~~~
\ha_n=\begin{pmatrix}a_n^+\cr a_n^-\end{pmatrix},~~~~~~
\hj=\begin{pmatrix}j^+\cr j^-\end{pmatrix}.
\ee
In Eq. \eqref{wpint}, the CTP inverse-propagators
\be
\hG^{-1}_{0n}=\begin{pmatrix}i\partial\br-m_n+i\epsilon&0\cr
0&-\gamma^0(i\partial\br-m_n+i\epsilon)^\dagger\gamma^0\end{pmatrix}+\hG^{-1}_{BCn},~~~~
\hD^{-1}_0 = \begin{pmatrix}\Box T+\xi\Box L+i\epsilon&0\cr0&-\Box T-\xi\Box L+i\epsilon\end{pmatrix}+\hD^{-1}_{BC},
\ee
already take into account boundary conditions in time, represented by the full
CTP matrices $\hG^{-1}_{BC}$ and $\hD^{-1}_{BC}$. The notation
\be
T^{ab}=g^{ab}-L^{ab},~~~L^{ab}=\frac{\partial^a\partial^b}{\Box}
\ee
will be used for the transverse and longitudinal projection operators.
The counterterm $S_{CT}$ will be suppressed below because it is given by 
the usual transition amplitude formalism.

The CTP propagators are introduced in Appendix \ref{ctppropa} for free fields. 
The bosonic two-point functions, such as the photon propagator, possess the structure
\be\label{scnf}
\hD = \begin{pmatrix}D&D^{+-}\cr D^{-+}&D^{--}\end{pmatrix}
=\begin{pmatrix}D^n&-D^f\cr D^f&-D^n\end{pmatrix}+iD_i\begin{pmatrix}1&1\cr1&1\end{pmatrix}
\ee
involving three real space-time dependent functions. It is an important 
property of radiative corrections that the self energies appearing in the 
quadratic generator functionals display the same structure and allows us to 
define the interactive near and far field propagators. The actual calculation 
of the propagator for photons reveals that $D^n$ and $D^f$ are the near and 
far field Green functions of classical electrodynamics, respectively. These 
propagators are usually introduced by splitting the Li\'enard-Wiechert potential 
of classical electrodynamics into an $\ord{r^{-1}}$ part and an $\ord{r^{-2}}$ 
part and the retarded and advanced propagators are defined as 
$D^{\stackrel{r}{a}}=D^n\pm D^f$. The propagator $\hD$ is symmetric with 
respect to the exchange of its indices, cf. Eq. \eq{freephotctp}. Therefore, 
the near and far field propagators can be identified by the symmetric and 
antisymmetric part of the retarded Green function, $D^n_{ab}=D^n_{ba}$, 
$D^f_{ab}=-D^f_{ba}$. Another, more microscopic separation of these 
two Green functions, based on the quantum level, is mentioned at the 
end of this section. Finally, we shall find in Section \ref{varprinc} an 
additional, classical way to define the near and far field propagators.

\subsection{Expectation values of the EM field and electric current}
Notice that the unitarity of the time evolution assures that expectation values
can be calculated in two different ways, e.g.
\be\label{expvatww}
\la A\ra=\fd{W}{j^+}_{|\hj=\ha=0}=\fd{W}{j^-}_{|\hj=\ha=0}
\ee
and the final time $t_f$ plays no role as long as the expectation value
is considered before it, $x^0<t_f$, in the equations above. Such a noncompact
$R$ symmetry, called the CTP symmetry, suggests the re-parametrization \cite{pol}
\be\label{jpar}
a^\pm=\frac{a}{2}(1\pm\kappa)\pm\ab,~~~~~~j^\pm=\frac{j}{2}(1\pm\kappa)\pm\jb
\ee
of the external sources because we have
\be\label{kapav}
\la A\ra=\fd{W}{j}_{|\hj=\ha=0}
\ee
for an arbitrary real value of $\kappa$. The book-keeping sources $a$ and $j$ must be 
canceled after calculating the functional derivatives in order to recover the physical
expectation values. On the contrary, the sources $\ab$ and $\jb$ can be nonvanishing
for unitary time evolution and represent the devices which are supposed to drive
the system adiabatically from the vacuum at $t=-\infty$ to the desired initial state
at $t=t_i$. Eq. \eq{kapav} shows that the fields coupled to $a$ and $j$ can be
identified by the physical fields because their averages reproduce the
field expectation values. We shall see later that the field coupled to
the sources $\ab$ and $\jb$ will be responsible for decoherence. 

The obvious difficulty of the construction of variational equations for the expectation 
values is the fact that we have twice as many degrees of freedom in the CTP
formalism as in the classical counterpart. This problem can be overcome
by considering the effective dynamics obtained by the elimination of a degree of 
freedom by its equation of motion from each CTP doublet. But the preceding argument
reveals an apparent problem in this plan. On the one hand, the reduplication of 
the degrees of freedom in the CTP formalism, $A\to A^\pm$, produces two 
identical looking variables as far as expectation values are concerned and it is
not clear which one to eliminate. On the other hand, decoherence, a mechanism not directly 
encoded by expectation values is related to the difference $A^d=A^+-A^-$ only. There
are different dynamical issues mixed in these variables. Therefore, it
appears to be appropriate to go over the Keldysh basis,
$(A^+,A^-)\to(A,A^d)=((A^++A^-)/2,A^+-A^-)$ and eliminate $A^d$. But there is a new
problem which arises in this manner. In fact, it is not difficult to see that the 
dynamics of the field variable $A$ and its canonically conjugated momentum 
$\Pi=-i\fd{}{A}$ is contained in the dependence of $A$ or $A^d$, respectively 
of the density matrix $\rho[A^+,A^-]$. The complete elimination of $A^d$
removes all information about the momentum in the effective theory. But the
time dependence of the expectation value $\la A\ra$ can be used to reconstruct
the expectation value $\la\Pi\ra=\partial_0\la A\ra$. Thus $A^d$ should appear
in the combination of $A^+$ and $A^-$ to be retained in the effective theory.
This condition is realized by the restriction $\kappa\ne0$ on the otherwise free parameter  \cite{pol}. 

The effective action, a device used to derive the equation of motion for the
expectation values, is usually the functional Legendre transform of $W$. This step must be
slightly modified because the latter object is complex in the double-time formalism. Since $W=0$
for the physical case with unitary time evolution we define the effective action as the 
Legendre transform of the real part of $W$,
\be\label{legtr}
\Re W[\ha,\hj]=\Gamma[\hJ,\hA]+\ha\hJ+\hj\hA,
\ee
and
\be\label{stof}
\hJ=\fd{W}{\ha},~~~~~~\hA=\fd{W}{\hj},
\ee
by using the external sources $a^\pm$ and $j^\pm$. The inverse transformation, Eq. \eq{legtr} and
\be\label{ftos}
\ha=-\fd{\Gamma}{\hJ},~~~~~~\hj=-\fd{\Gamma}{\hA},
\ee
appear as variational equations of motion satisfied by the fields $\hJ$ and $\hA$. It is
advantageous to find the effective action for the expectation values. To this end we write
\be\label{effaphav}
\Gamma[J,\Jb,A,\Ab]=\Re W[\ha,\hj]-aJ-\ab\Jb-jA-\jb\Ab
\ee
by means of the parametrization \eq{jpar} and
\be\label{phauxvar}
J=\fd{\Re W}{a},~~~~~~\Jb=\fd{\Re W}{\ab},~~~~~~
A=\fd{\Re W}{j},~~~~~~\Ab=\fd{\Re W}{\jb}.
\ee
The inverse transformation,
\be
a=-\fd{\Gamma}{J},~~~~~~\ab=-\fd{\Gamma}{\Jb},~~~~~~j=\fd{\Gamma}{A},~~~~~~\jb=\fd{\Gamma}{\Ab},
\ee
represents the Euler-Lagrange equations. The fields $J$ and $A$ stand for the expectation values 
and the auxiliary fields $\Jb$ and $\Ab$ incorporate the effects of quantum fluctuations.

The imaginary part $\Im W$, left out from the construction of the our effective action
is important in establishing decoherence. In fact, 
decoherence stands for the suppression of the off diagonal elements of 
the density matrix in the pointer representation. By assuming that
the field variables $J$ and $A$ are good pointer variables decoherence corresponds
to the suppression of the absolute magnitude of $e^{iW[\ha,\hj]}$ as $\Jb$ and $\Ab$ are
increased. Such a suppression comes from $\Im W[\ha,\hj]$.

Finally, one can construct an effective action for the physical expectation values only,
\be
\Gamma[J,A]=\Re W[a,\ab,j,\jb]-aJ-jA
\ee
where
\be
J=\fd{\Re W}{a},~~~~~~A=\fd{\Re W}{j}.
\ee
The inverse transformation,
\be
a=-\fd{\Gamma}{J},~~~~~~j=-\fd{\Gamma}{A},
\ee
give the Euler-Lagrange equations for the expectation values only.

\subsection{Reduced density matrix}
We have surveyed so far the means to find equations of motion for expectation values.
The right-hand side of Eq. \eq{expv} is not sufficient for our other goal, for the calculation of the
reduced density matrix. The reduced density matrix for the EM field is obtained by 
eliminating the charged degrees of freedom and can be written as
\be\label{densm}
\rho[A^{(1)},A^{(2)}]=\la A^{(1)}|\Tr_{ch}[Ae^{-i(t-t_i)H}\rho_ie^{i(t-t_i)H}]|A^{(2)}\ra
\ee
where the trace operation is over the Fock space of the charged particles and the state
$|A\ra$ is the eigenstate of the photon field operator with field configurations
$A_\mu(\v{x})$ in the functional Schr\"odinger representation. The generator functional of 
Eq. \eq{wpint} for $\rho[A^{(1)},A^{(2)}]$
will be given in the OTP formalism. It consists of a path integral 
where the charged fields have a closed time path and the integration over the photon field is 
restricted to trajectories with open end points $A^+_\mu(t_f,\v{x})=A_\mu^{(1)}(\v{x})$ and
$A^-_\mu(t_f,\v{x})=A_\mu^{(2)}(\v{x})$. 

The OTP formalism displays entanglement of subsystems in a specially clear manner.
Let us consider for instance QED as a closed system, described by a factorizable, pure state
at the initial time. The interactions  generate entanglement between the charges 
and the EM field; the charges considered alone will be found in a 
mixed state and their reduced density matrix can be obtained by eliminating the 
EM field. The contributions of the perturbative integration over the EM field 
configurations are labeled by graphs. Those which contain photon lines
connecting the two time axes represent mixed state contributions. In fact,
the integration over the momentum of these lines produces the sum of pure state
contributions to the reduced density matrix. The structure \eq{scnf} of the photon
propagator reveals that the near field is responsible for the interactions
within a time axis. These elementary processes contribute to the self energy 
and preserve the purity of the charge states. The far field connects the two 
time axes and represents the mixing resulting from the entanglement of the
charge-EM field subsystems.

The perturbative evaluation of the generator functional of Eq. \eq{waj} implies a strong 
limitation. The naive perturbation expansion relies on the translation invariance of the 
vacuum, allowing weak inhomogeneous components in the external sources. This is
enough to support sufficiently inhomogeneous field expectation value profiles
when there is no gap in the excitation spectrum above the vacuum, such as in the
case for a system of charges at finite density. But once the localized states
arise through a gap in the spectrum then the weak external inhomogeneities
are not sufficient to produce localized states on the mass-shell. In fact,
all we achieve at finite temperature and vanishing charge density is a weak 
inhomogeneous vacuum polarization by virtual pairs. The creation of states
with few on-shell localized charges remains a nonperturbative issue in this 
formalism \cite{pol}.

\section{Fluctuations and dressing}\label{fldres}
Both the external, classical source $\hj$ and the dynamical current $\la0|\hj|0\ra=\hJ$
generate vacuum polarization. The simple structure of the QED action allows us to 
compare these two polarizations. To this end we consider the expectation value of 
the EM field induced by weak external and dynamical, quantum currents. We 
shall seek the linearized expression for the EM field. The result will be equally valid 
for the transition amplitude formalism or for any double-time functionals. The 
integration over the fermion fields in the generator functional \eq{wpint} yields
\be\label{whahj}
e^{iW[\ha,\hj]}=\int D[\hA]e^{iW^c[\ha-e\hsigma\hA]+\frac{i}{2}\hA\hD_0^{-1}\hA+i\hj \hA}
\ee
where $\hsigma$ denotes flipping the sign of the $-$ component of CTP doublets
\be
\hsigma\begin{pmatrix}A^+\cr A^-\end{pmatrix}=\begin{pmatrix}A^+\cr-A^-\end{pmatrix},
\ee
the generator functional of non-interacting fermions is denoted by
\be\label{frefgf}
W^c[\ha]=\sum_n W_n[\ha_n],
\ee
with
\be\label{frefgfk}
W_n[\ha]=-i\Tr\ln\left[\hG^{-1}_{0n}+\begin{pmatrix}a\br^+&0\cr0&a\br^-\end{pmatrix}\right]
\ee
being the generator functional for the connected Green functions of the current for the
flavor $n$ in the absence of electromagnetic interactions. Note that there are
connected Green functions of arbitrarily high order for the noninteracting Dirac-see 
because the current is a composite operator. We perform the shift of the integral variable, 
$\hA\to \hA-\hD_0\hj$, in other words, we use the equation of motion for the EM field to write
\bea\label{intf}
e^{iW[\ha,\hj]}&=&e^{-\frac{i}{2}\hj\hD_0\hj}\int D[\hA]e^{iW^c[\ha-e\hsigma(\hA-\hD_0\hj)]+\frac{i}{2}\hA\hD_0^{-1}\hA}\nn
&=&e^{iW[\ha+e\hsigma\hD_0\hj]-\frac{i}2\hj\hD_0\hj},
\eea
with
\be\label{wacsak}
W[\ha]=W[\ha,\hj=0]
\ee
being the generator functional for the current in full QED. Therefore, we obtain the relation
\be\label{wsimpl}
W[\ha,\hj]=W[\ha+e\hsigma\hD_0\hj]-\hf\hj\hD_0\hj,
\ee
the reduction of the two-variable generator functional into a single variable functional.
The simple form of the initial Lagrangian allows us to separate the tree-level
dependence on the source $j$ and to place the loop-induced dependence into the 
dressed fermion one-loop graphs. In doing so, we rely on the fact that the only part 
of the QED action which contains higher-than-quadratic terms in the fields is 
the minimal coupling.

The natural use of such a reduction is the simplification of the equations of motion for the currents
\be\label{elcurref}
\hJ_n=\fd{W[\ha+e\hsigma\hD_0\hj]}{\ha_n}
\ee
and the EM field
\be\label{evaexe}
\hA=\sum_n e_n\hD_0\hsigma\fd{W[\ha+e\hsigma\hD_0\hj]}{\ha_n}-\hD_0\hj
\ee
which can be written as
\be\label{evaex}
\hA=\hD_0(\hsigma e^{tr}\hJ-\hj)
\ee
where $e$ denotes a flavor column vector made up of the electric charges, $e^{tr}=(e_0,e_1,\cdots)$.

The external sources $\ha_n$ and $\hj$ are the independent variables in these equations.
When the functional Legendre transformation is performed then $\hJ_n$ and $\hA$ become
independent variables but the equations remain valid by considering $\hj$ and $\ha_n$ as dependent variables.
Comparing Eq. \eq{evaex} with the second equation in Eqs. \eq{ftos} the exact result \cite{kornel}
\be\label{effactnl}
\Gamma[\hJ,\hA]=\Gamma_{mech}[\hJ]+\hf\hA\hD_0^{-1}\hA-\hA\hsigma\sum_n e_n\hJ_n
\ee
follows. Here the first term represents the mechanical contribution of the Dirac-see and
plays the role of an integration "constant". There are two sources of complexities in the 
dynamics, the implications of the Pauli exclusion principle for the electric current as 
a composite operator and the electromagnetic interactions. The latter has a small parameter, 
$\hbar$ or $e^2$ to organize a systematical approximation scheme but the former type of 
dynamical correlations admit no small parameter and induce arbitrary high order 
connected Green functions for the non-interacting Dirac-see in Eq. \eq{frefgfk}
and arbitrary high order vertices in the effective action \eq{effactnl}. 

The lesson of Eq. \eq{evaex} is rather surprising, it is the absence of 
renormalization in the nonmechanical part of the theory. We shall see below,
when a similar result is recovered in the loop-expansion, that this triviality results 
from a cancellation between the two channels the source $j$ can induce
EM field. What is observed here is that the relation among the expectation values and the
external current $j$ is not renormalized because the former already includes all the dressing. 
The nontrivial source of dressing according to Eq. \eq{wsimpl} arises
from the Green functions for the currents, the structure of the functional $W[\ha]$ of 
Eq. \eq{wacsak}. The $j$ dependence comes in a trivial manner, dictated by the minimal 
coupling, gauge invariance. 

It is instructive to follow what happens with the vacuum polarization when we select 
the flavor $n=0$ as our valence charge, $e_v=e_0$, and cancel the external source $a_{n_b}$
$n_b\not=0$ for the remaining flavors, called background charges. Let us
introduce the notation $\ha_v=\ha_0$,
\be\label{valcur}
\hJ_v=\fd{W[\ha+e\hsigma\hD_0\hj]}{\ha_v},
\ee
and $e_b$ will denote the background charge vector, $e_b^{tr}=(0,e_1,\cdots)$. 
The polarization effects now arise from the dependence of the electric current of any flavor on the
external sources $\ha_v$ and $\hj$ and one can find these contributions by expanding Eq. \eq{elcurref}
in $\ha_v$ and in $\hj$. We have to keep in mind that the current of the valence charge, given 
by Eq. \eq{valcur}, includes all external source dependence and it is an independent variables 
of the effective action therefore, its dependence on $j$ has already been accounted for. Thus 
the expansion can be restricted to the background charges which yields, after inserting it into Eq. \eq{evaexe},
\be\label{maxeqvc}
\hA=\hD_0\left[\hsigma e_v\hJ_v-\hj+\hsigma e^{tr}_b\fdd{W[\ha]}{\ha}{\ha_v}_{|\ha=0}\ha_v
+\hsigma e^{tr}_b\fdd{W[\ha]}{\ha}{\ha}_{|\ha=0}e_b\hsigma\hD_0\hj\right]+\cdots
\ee
up to quadratic terms in the external sources for charge conjugation invariant vacuum.
The first two terms on the right-hand side stand for the direct, tree-level part of Eq. \eq{evaex}
and the polarization effects are represented by the third and fourth terms in the linearized equation 
of motion. Recall that the external field $\ha_v$ is used only to generate the valence current $\hJ_v$.
Therefore, $\ha_v$ in the third term should be expressed in terms of $\hJ_v$ when the relation between 
the valence current and the EM field is sought. Thus this term represents the polarization
effects due to the dependence of the background currents on the valence current. The last
term shows the polarization due to the dependence of the background currents on the 
external current $\hj$. These two terms are not directly incorporated in the valence current 
dynamics and they represent the dressing in the equation of motion when $\ha_v$ is expressed
in terms of $\hJ_v$. In other words, the dressing in the Maxwell equation \eq{maxeqvc}
is due to the fact that our experimentally monitored valence current does not cover all charges 
in the system. It is completely natural that the uncontrolled charges generate the 
dressing, which gives rise to the proportionality between the averaged and the local EM field.

\section{Effective action}\label{cgrefnct}
After some general remarks we now turn to explicit loop-expansion expressions for the 
quadratic part of the functionals mentioned in the previous Section.

\subsection{Connected Green functions}
The generator functionals of Eq. \eq{whahj} for a single valence charge $n=0$ can be written after 
carrying out the integration over the fermion fields as
\be\label{whahjfe}
e^{iW[\ha,\hj]}=\int D[\hA]e^{\sum_n W_n[\delta_{0,n}\ha]+i\hj\hA
+\frac{i}{2}\hA\hD^{-1}\hA+\ord{\ha^2}+\ord{\hA^3}}
\ee
where $W_n$ is defined by Eq. \eq{frefgfk} and a homogeneous, classical 
background charge is assumed to cancel the
$\ord{\hA}$ tadpole terms when the vacuum is charged. The improved photon propagator
\be\label{fullphpr}
\hD=(\hD^{-1}_0-\hat\Pi)^{-1}
\ee
contains the one-loop polarization from all charges,
\be\label{totse}
\hat\Pi=\hsigma e^{tr}\htG e\hsigma,
\ee
where
\be
\tG_{(n\sigma x\mu),(n'\sigma' y\nu)}=-i\delta_{nn'}\tr(G^{\sigma'\sigma}_{0n~yx}
\gamma^\sigma_\mu G^{\sigma\sigma'}_{0n~xy}\gamma^{\sigma'}_\nu).
\ee
The background charge improved photon propagator, used later is given by 
\be\label{db}
\hD_b=\frac1{\hD^{-1}_0-\hat\Pi_b}
\ee
including the polarization
\be\label{pib}
\hat\Pi_b=\hsigma e^{tr}_b\htG e_b\hsigma.
\ee

The CTP expression for the functional $W[\ha,\hj]$ is obtained when the photon field trajectories
are closed in the functional integral of Eq. \eq{whahjfe}. We shall need the OTP result which
is obtained by carrying out this functional integral for uncorrelated $A^\pm$
trajectories, having different end points. The functional integration yields
\be\label{quwctp}
W[\ha,\hj]=-\hf(\ha,\hj)\begin{pmatrix}\htG_v&\htG_e\hsigma\hD_0\cr
\hD_0\hsigma\htG_e&\hD_t\end{pmatrix}\begin{pmatrix}\ha\cr\hj\end{pmatrix}
+\ord{\hbar^2}+\ord{\mr{source}^3}
\ee
where $\htG_v=\htG_{00}$, $\htG_e=(\htG e)_0$. 

The dressed propagators are the sum of the products $\hD_1\hsigma\hD_2\hsigma\cdots\hsigma\hD_n$ 
where $\hD_j$ has the form shown in Eq. \eq{scnf}. The product displays the 
same structure as the factors $\hD_j$, namely the retarded or advanced part of the 
product is the product of the retarded or advanced parts,
\be\label{prod}
(\hD_1\hsigma\hD_2\hsigma\cdots\hsigma\hD_n\hsigma)^{\stackrel{r}{a}}
=D_1^{\stackrel{r}{a}}D_2^{\stackrel{r}{a}}\cdots D_n^{\stackrel{r}{a}},
\ee
yielding
\be\label{dbr}
D^{\stackrel{r}{a}}=\frac1{D^{\stackrel{r}{a}-1}_0-\Pi^{\stackrel{r}{a}}}.
\ee

We need the real part of $W$, considered in real space as opposed to momentum space,
for the effective action in Eq. \eq{effaphav}. To simplify matters 
we exclude pair creation processes by restricting ourselves external sources 
with modes $\omega,k<m$ and the heat and particle baths are chosen to be 
nonrelativistic, $T,k_F\ll m$. The one-loop expressions for both $\Re\htG$ and 
$\Im\htG$ are vanishing on the photon mass-shell, the support of $\Im D_0$,
according to the calculation reported in Section \ref{olplta}. These properties 
allows us to simplify the real part of the products in the matrix elements in
Eq. \eq{quwctp} and one finds
\be\label{rew}
\Re W[\ha,\hj] = -\hf(\ha,\hj)\begin{pmatrix}\Re\htG_v &e_v\Re(\htG_v)\hsigma\Re(\hD_{0})\cr
e_v\Re(\hD_{0})\hsigma\Re(\htG_v)&\Re\hD_{0}+\Re(\hD_{0})\Re(\hat\Pi)\Re(D_{0})\end{pmatrix}
\begin{pmatrix}\ha\cr\hj\end{pmatrix},
\ee
The form \eq{scnf} of $\htG$ and $\hD$ gives after a lengthy but straightforward calculation
\be\label{wallf}
\Re W[\ha,\hj]=-\hf^{(a,j,\ab,\jb)}\begin{pmatrix}-\kappa\tG^n&\kappa e_v\tG^nD^n_0&-\tG^r&e_v\tG^rD^r_0\cr
\kappa e_vD_0^n\tG^n&-\kappa D^n&e_vD^r_0\tG^r&-D^r\cr-\tG^a&e_v\tG^aD^a_0&0&0\cr
e_vD^a_0\tG^a&-D^a&0&0\end{pmatrix}\begin{pmatrix}a\cr j\cr\ab\cr\jb\end{pmatrix}
\ee
where $\tG^{\stackrel{r}{a}}$ denotes $(\Re\tG)^{\stackrel{r}{a}}$.
This result leads to the linearized expressions
\bea\label{expvitst}
J&=&\kappa\tG^na-\kappa e_v\tG^nD_0^nj+\tG^r\ab-e_v\tG^rD_0^r\jb\nn
A&=&-\kappa e_vD_0\tG^na+\kappa D^nj-e_vD_0^r\tG^r\ab+D^r\jb\nn
J^a&=&\tG^aa-e_v\tG^aD_0^aj\nn
A^a&=&-e_vD_0^a\tG^aa+D^aj
\eea
for the expectation values in terms of the external sources for all charges or for a 
single valence charge, respectively. 

Notice that the physical expectation values $J$ and $A$ 
contain the retarded field of the physical sources $\ab$ and $\jb$. This is not a dynamical
breakdown of the time reversal invariance, but rather a trivial result of the boundary conditions.
In fact, the open ended, null boundary condition at the final time leads to a
destructive interference between the two time axes which cancels the advanced field 
generated by the physical sources \cite{pol}. The book-keeping variables $a$ and $j$ generate 
an $\ord{\kappa}$ time reversal invariant near field, as a result of the equal coupling of 
these sources to the dynamical variables of the two time axes by the $\kappa$-dependent term 
in Eq. \eq{jpar}. 

After the formal manipulations we set $a=j=0$ to regain the physical case 
with unitary time evolution. The auxiliary fields are vanishing in this case, cf. Eq. \eq{expvatww},
and the Maxwell equation, the expression of the EM field in terms of the dressed, 
retarded analogy of the Li\'enard-Wiecher potential, reads 
\be\label{maxallp}
A=D^r_b\jb-e_vD_0^rJ
\ee
where $D^{\stackrel{r}{a}-1}_b=D^{\stackrel{r}{a}-1}_0-\Pi^{\stackrel{r}{a}}_b$.

\subsection{Legendre transform}
The variation principle governing the expectation values is based on the effective
action obtained by a functional Legendre transformation of $\Re W[\ha,\hj]$.
We mention first a peculiar feature of the loop-expansion. 
The independent variables of the generator functional for the connected Green 
functions $W[\ha,\hj]$ are the external sources $\ha$ and $\hj$ which are classical, 
$\ord{\hbar^0}$ quantities. The result is the well-known equivalence of the 
expansion in $\hbar$ and in the number of loops in the Feynman graphs in the 
connected Green functions. But in the present case the electric current $\hJ$ 
is $\ord{\hbar}$ and the expansion in $\hbar$ mixes different loop-orders in the effective action. 

The Legendre transform of a quadratic functional
remains quadratic, and the two kernels are the inverse of each other up to a sign.
The inverse of the block matrices of the quadratic functionals of Eq. \eq{quwctp}
can be obtained by performing the change of variable of the Legendre transformation 
explicitly with the result
\be\label{eqGamma}
\Gamma=-\hf(J,A,J^a,A^a)\begin{pmatrix}0&0&\tG^{a-1}&e_v\cr0&0&e_v&D_b^{a-1}\cr
\tG^{r-1}&e_v&-\kappa\tG^{r-1}\tG^n\tG^{a-1}&-\kappa(P_{off}+\tG^{r-1}\tG^nP_{on})e_v\cr
e_v&D^{r-1}_b&-\kappa e_v(P_{on}\tG^n\tG^{a-1}+P_{off})&-\kappa(\Pi^n_bP_{on}+D_b^{n-1}P_{off})\end{pmatrix}
\begin{pmatrix}J\cr A\cr J^a\cr A^a\end{pmatrix},
\ee

The operator $P_{on}$ projects on to the photon mass-shell and $P_{off}=\openone-P_{on}$. Some important
properties of $D^n$ and $D^f$ related to the mass-shell and used in the derivation
are listed in Table \ref{nfprpr}. For instance, the replacements
$D^{r-1}_0D^n_1\Pi_t^n(\openone-D_0^nD_0^{a-1})\to0$,
$D^{r-1}_0D_0^nD_0^{a-1}\to D_b^{n-1}P_{off}$ and 
$\Pi^n_b(\openone-D^n_0D^{a-1}_0)\to\Pi^n_bP_{on}$
have been carried out in obtaining the $\ord{A^a}$ term in the effective action.
The corresponding linearized, one-loop equations of motion are
\bea\label{eqmafa}
a&=&e_vA^a+\tG^{a-1}J^a\nn
j&=&D^{a-1}_bA^a+e_vJ^a\nn
\ab&=&\tG^{r-1}J-\kappa\tG^{r-1}\tG^n\tG^{a-1}J^a+eA-\kappa(P_{off}+\tG^{r-1}\tG^nP_{on})e_vA^a\nn
\jb&=&D^{r-1}_bA-\kappa(\Pi^n_bP_{on}+D_b^{n-1}P_{off})A^a-\kappa e_v(P_{on}\tG^n\tG^{a-1}+P_{off})J^a+e_vJ
\eea
Notice that the auxiliary fields couple in a different manner to the physical external sources 
on and off the mass shell. 
The last expression is the vacuum polarization corrected Maxwell equation. It corresponds to 
the equation of motion Eq. (4.18) in Ref. \cite{cooper} when the effects of the physical 
sources which drive the system adiabatically to the desired initial state are incorporated 
implicitly in the propagators and the self-energy.

\begin{table}
\caption{Properties of the near and far propagator on and off the mass-shell}\label{nfprpr}
\begin{ruledtabular}
\begin{tabular}{lcr}
Operator&Mass-shell&Off mass-shell\\
\hline
$\Box D^n$&0&1\\
$D_0^n$&0&$-\Box^{-1}$\\
$D_0^f$&$\neq0$&$0$\\
$D_0^nD_0^{\stackrel{r}{a}-1}$&0&$1$\\
$D_0^{\stackrel{a}{r}-1}D_0^nD_0^{\stackrel{r}{a}-1}$&0&$D_0^{n-1}$\\
\end{tabular}
\end{ruledtabular}
\end{table}

The auxiliary fields are not needed for the dynamics of the observable expectation values 
therefore, it is natural to seek the effective action involving the physical fields alone. 
To this end we consider $\ab$ and $\jb$ as parameters in the generator functional of 
Eq. \eq{wallf} and perform the functional Legendre transformation on $a$ and $j$ only, 
with the result
\be\label{effaphf}
\kappa\Gamma[J,A]=\hf^{(J,A)}\begin{pmatrix}\tG^{n-1}&-e_v\cr-e_v&D^{n-1}\end{pmatrix}
\begin{pmatrix}J\cr A\end{pmatrix}+Ja_v''+Aj'',
\ee
where the inverse of the near field propagators is defined to be zero on the mass-shell.
One may even keep only one field performing the Legendre transformation for only a single source 
with the result
\be\label{effaa}
\kappa\Gamma[A]=\hf AD^{n-1}_tA+Aj',
\ee
and
\be
\kappa\Gamma[J]=\hf J\tG^{n-1}J+Ja'.
\ee
The source terms $a'',j'',a'$ and $j'$ are the sum of the action of different retarded propagators
on $\ab$ and $\jb$, their detailed form is such that Eqs. \eq{expvitst}
are satisfied. Notice the need for $\kappa\not=0$ to arrive at any variational scheme
for the physical variables only \cite{pol}. 

What is the scale regime where the renormalization group treatment, performed
in its complexity is supposed to preserve the form of these equations? 
The UV cutoff of the equations is the Compton wavelength because the vacuum polarization effects 
suppress the variation in the space-time within this distance scale.
The infrared limitation is provided by the typical length scale of the
state generated by the external sources, such as the de Broglie wave length. In 
fact, the smearing of the configuration beyond this length scale modifies the 
space-time dependence and the equations of motion. Once the 
blocking scale passes the quantum-classical crossover one arrives at the
true, classical equations supported by decoherence.

\subsection{Radiation field}\label{varprinc}
The external sources or boundary conditions in time determine the radiation field in 
classical electrodynamics. The non-trivial issue here is the realization of this
well-understood circumstance within the variational principle. The point is that the
radiation field enters, in a singular manner, into the variational scheme of the 
actions \eq{effaphf}-\eq{effaa}. The first sign of this complication is the vanishing 
of the near field  propagator on the photon mass-shell due to the principal 
value prescription in its Fourier integral representation. Another way to see that the
radiation field does not enter into the quadratic part of the action is to recall the
remark made after Eq. \eq{scnf}, namely that the far field propagator is antisymmetric
thus the radiation field drops out entirely from the $\ord{A^2}$ part of the action. 
This state of affairs is in agreement with the formal time reversal symmetry of the
elementary processes because the field operators act forward and backward in 
time by annihilating and creating elementary excitations. 

The retarded propagators may appear in the $\ord{Aj'}$ terms of the action and are
attached to the external sources. The $\ord{A^2}$ part of the action controls
the restoring force acting on the oscillation around the stable vacuum thus 
the external source induces a singular response on the mass shell, rendering the
mass-shell radiation field nondynamical, to be settled by the boundary conditions in time.
In fact, a simple way to arrive at the desired equations of motion
without separating explicitly the on- and off-shell components of the equation as 
in Eqs. \eq{eqmafa} is to write the source in Eq. \eq{effaa} as 
$j'=D^{n-1}A_{on}$ and note that the equation of motion $A=A_{on}$ 
contains the mass-shell component with a formal 0/0 coefficient. This scheme is
actually realized by the usual $m^2\to m^2-i\epsilon$ prescription for 
the free action when applied in classical field theory.

When all fields arising from the reduplication of the degrees of freedom inherent 
in the double-time formalism are retained then the physical fields can be coupled to 
nonsymmetrical kernels. The result is the simple and natural preservation of the retarded 
and advanced solutions in the variational scheme. Such a treatment of the boundary 
conditions in time opens up possibilities to address the genuine, dynamical breakdown of time 
reversal invariance. For instance, the friction forces signal in classical physics that the 
coupling of our system to its environment breaks the time reversal invariance in a 
dynamical manner. Such kind of forces can easily be encoded in the variation 
principle of the double-time formalism. Another example is the dynamical building 
up of decoherence in a system due to a coupling to its environment with a gapless spectrum,
to be seen explicitly below.

\section{One-loop polarization tensor}\label{olplta}
We return in this Section to the loop expansion and present the photon-polarization tensor
in the one-loop approximation. We shall work with the Fourier transform
\be
\tG^{\sigma\sigma'}_{(x\mu)(y\nu)}=\int_qe^{-ixq}\tG^{\sigma\sigma'}_{q\mu\nu},
\ee
where the integration in the Fourier space is denoted by
\be
\int_p=\int\frac{d^dp}{(2\pi)^d}.
\ee
The real and imaginary parts of the two-point function are defined in the space-time. 
Their Fourier transform is
\be
(\Re\htG)_q=\hf(\htG_q+\htG^*_{-q}),~~~(\Im\htG)_q=\frac1{2i}(\htG_q-\htG^*_{-q}).
\ee
Our goal is to obtain 
\be\label{onelpoltctp}
\tG^{(\sigma\mu)(\tau\nu)}_q=-\frac{i\hbar}4\sum_{\eta,\eta'}\int_p
\tr[\gamma^\mu G^{\sigma \tau}_{q+p}\gamma^\nu G^{\tau\sigma}_p]
\ee
for a single charge in terms of the CTP propagator \eq{fermionctp}. Because of the structure
\eq{scnf} it is sufficient to find $\tG^{++}$ and $\tG^{+-}$. It will
be useful to recall that this satisfies the Ward identity 
\be\label{ward}
k^\mu\htG_{\mu\nu}=0.
\ee
For the sake of simplicity we seek the detailed expressions below the pair 
creation threshold, $q^2<4m^2$, and keep the density and the temperature low
enough to keep the charges nonrelativistic. 

The CTP propagators are the sum of the contributions from the vacuum and from
the environment. The latter is linear in the Fermi-Dirac distribution function 
$n_p=1/\{\exp(\beta[\epsilon_\v{p}-\mr{sign}(p^0)\mu])+1\}$.  
As a result we have the similar separation in the polarization tensor
\eq{onelpoltctp}, which will be the sum of 
three kinds of terms. The vacuum contributions are independent of $n$. 
The $\ord{n}$ terms contain a particle from the environment on the mass-shell.
Finally, there will be $\ord{n^2}$ terms where both particles come from the 
environment and are on the mass-shell.

As far as the environment is concerned, we shall have either finite temperature at 
vanishing density or finite density at vanishing temperature. We write the two-point
function as $\htG=\htG_{vac}+\htG_{env}$ where $\htG_{env}$ depends on the 
temperature $T$ or the Fermi momentum $k_F$. The matrix element $++$ of the 
polarization tensor in the vacuum is well known, its form is
\be\label{ppv}
\tG^{++\mu\nu}_{vac,q}=\frac{\alpha}{3\pi}q^2\left(g^{\mu\nu}-\frac{q^\mu q^\nu}{q^2}\right)
\left\{\frac13+2\left(1+\frac{2m^2}{q^2}\right)
\left[\sqrt{\frac{4m^2}{q^2}-1}~\mr{arccot}\sqrt{\frac{4m^2}{q^2}-1}-1\right]\right\}.
\ee
with $\alpha=e^2/4\pi$. The matrix element $+-$,
\be\label{pmv}
\tG^{+-\mu\nu}_{vac,q}=-128\pi^3i\alpha m^2\left(g^{\mu\nu}-\frac{q^\mu q^\nu}{q^2}\right)
\left(1+\frac{q^2}{2m^2}\right)\int_p\delta(q^2+2pq)\delta(p^2-m^2)\Theta(-p^0-q^0)\Theta(p^0)
\ee
is vanishing below the pair creation threshold because it consists of on-shell amplitudes only.

The environment breaks the formal Lorentz covariance of $\htG_{env}$. Let us denote the temporal
unit vector by $u^\mu$ which assumes the form $u^\mu=(1,\v{0})$ in the rest frame of the 
environment where the expressions for the distribution functions of Eqs. 
\eq{distrfncs} are valid. We use two invariants $\v{q}^2=-[q-u(uq)]^2$ and 
$\nu=uq/|\v{q}|=\omega/|\v{q}|$ with $q^\mu=(\omega,\v{q})$ to parameterize the 
momentum dependence. For each CTP index, covariance with respect to three-dimensional rotations 
suggests that we use three three-dimensional scalars whose number is further reduced 
to two by gauge invariance. These two sets of scalar are the easiest to obtain in terms
of $\hat \TERMA =g_{\mu\nu}\htG^{\nu\mu}$ and $\hat \TERMB =u\htG u$. The final useful form is
\be\label{phse}
(\htG_q)^{\mu\nu}=\hat \TERMB_q\begin{pmatrix}1&\v{n}\nu\cr\v{n}\nu&\nu^2\v{L}\end{pmatrix}
+\hf[\hat \TERMB_q(1-\nu^2)-\hat \TERMA_q]\begin{pmatrix}0&0\cr0&\v{T}\end{pmatrix}
\ee
where the space-time tensor structure is explicitly shown for the environment contributions
with $\v{n}=\v{q}/|\v{q}|$, $\v{L}=\v{n}\otimes\v{n}$, $\v{T}=\openone-\v{L}$. 

The expression \eq{phse} makes the space-time tensor structure explicit. The CTP index 
structure is contained in the Lorentz-scalar CTP matrices $\hat \TERMA$ and $\hat \TERMB$ which will 
be detailed now. These matrices are the sum of the vacuum and environment contributions,
\be
\hat X_q=\hat X_{vac,q}+\hat X_{env,q}
\ee
where $X$ stands for $\TERMA$ or $\TERMB$. As far as the vacuum contribution is concerned
in the framework of the gradient expansion it contains $\tG^{++\mu\nu}_{vac,q}$ only which gives
\be\label{vachatab}
(\Re\hat \TERMB_{vac})_q=\frac{\alpha}{15\pi}\frac{q^2}{m^2}\v{q}^2\begin{pmatrix}1&0\cr0&-1\end{pmatrix}
=\frac{(\Re\hat \TERMA_{vac})_q}{3(1-\nu^2)}
\ee
and $\Im\hat \TERMA_{vac}=\Im\hat \TERMB_{vac}=0$ with the help of Eq. \eq{ppv}. The structure of Eq. \eq{scnf} 
is preserved by the environment contribution, and we find the real and imaginary parts
\bea\label{reimpa}
(\Re X^{++}_{env})_q&=&\Re\int_p2\pi\delta(p^2-m^2)P\frac{X_{p,q}(n_p+n_{-p})}{q^2+2pq+i\epsilon}\nn
(\Re X^{+-}_{env})_q&=&\frac{i}2\int_p2\pi\delta(q^2+2pq)2\pi\delta(p^2-m^2)
\mr{sign}(p^0+q^0)(n_p+n_{-p})X_{p,q}\nn
(\Im X_{env})_q&=&\hf\int_p2\pi\delta(q^2+2pq)2\pi\delta(p^2-m^2)
[n_p(n_{q+p}-1)+n_{-p}(n_{-p-q}-1)]X_{p,q},
\eea
where $P$ denotes the principal value integral and the kinematical factors are
\bea
\TERMA_{p,q}&=&8(m^2-pq),\nn
\TERMB_{p,q}&=&8\left(p^{02}+p^0q^0-\hf pq\right).
\eea
We shall consider the cases of finite temperature and vanishing density and
finite density and vanishing temperature when the one-particle distribution function 
$n_p$ generates the scales $k_{ch}=T$ and $k_{ch}=k_F$, respectively and 
keep the environment in the nonrelativistic regime, $T,k_F\ll m$ for simplicity.
The frequency integrals are carried out in Eqs. \eq{reimpa} 
by means of the residuum theorem, resulting in three-dimensional integration over the momentum 
$\v{p}$. By splitting these integrals into two parts, corresponding to $|\v{p}|<m$ and
$|\v{p}|>m$ and by finding an upper bound for the latter contributions
one can verify that the simple nonrelativistic expressions give reliable leading 
order approximations. 

What are the kinematical regions contributing to the various pieces of the
two-point function for a nonrelativistic environment? At low temperature and 
vanishing charge density it is enough to retain the 
$\ord{n}$ terms which decreases rapidly when $|\v{q}|$ is increased. The situation 
changes drastically at finite density because the gap disappears in the excitation 
spectrum. $\Re X^{++}$, given in the first equation of Eq. \eq{reimpa} 
describes a particle-hole pair where one member of the pair belongs to the 
environment and is on the mass shell. It is advantageous to use the dimensionless ratio
\be\label{rdef}
r=\frac{q^2+2\omega_pq^0}{2|\v{q}|k_{ch}}
\ee
which gives
\be\label{disprel}
\omega\approx\frac{(\v{q}+rk_{ch}\v{n})^2}{2m}-\frac{(rk_{ch})^2}{2m}
\ee
indicating that the Pauli blocking allows
particle-hole excitations at small $\v{q}$ for $|r|\approx1$. Thus $\Re X^{++}$ 
is expected to take larger values for $\v{q}\ll m$ at $|r|\approx1$ only.
Both the particle and the hole are on the mass-shell in the remaining expressions 
in Eqs. \eq{reimpa} because the time ordering is the only mechanism to produce
off-shell amplitudes in the CTP propagators. The distribution function 
$n_p$ restricts the integration approximately to $|\v{p}|<k_{cr}$. The conditions 
$p^2=(p-q)^2=m^2$ for nonrelativistic four momentum $p$ with $p^0\approx m$ give 
$rk_{cr}=\v{p}\v{n}$, the integration is over a plan, orthogonal to
$\v{q}$. The largest area corresponds to $r=0$ and the remaining functions in
Eqs. \eq{reimpa} tend to be large for $r\approx0$.

The actual calculation of the integrals leads to the expressions
\bea\label{reimx}
\Re X^{++}_q&=&X^+_{q^0,\v{q}}+X^+_{-q^0,\v{q}},\nn
\Re X^{+-}_q&=&i(X^-_{-q^0,\v{q}}-X^-_{q^0,\v{q}}),\nn
\Im X^{++}_q&=&X^i_{q^0,\v{q}}+X^i_{-q^0,\v{q}},
\eea
where the case of finite temperature and vanishing density gives
\bea\label{reimxt}
\TERMA^+_q&=&\frac{4\alpha m^2T}{\pi|\v{q}|}\left(1+\frac{q^2}{2m^2}\right)I_q-\frac{4\alpha T^2}{\pi}\tilde I_q,\nn
\TERMA^-_q&=&-\frac{4\alpha m^2T}{|\v{q}|}\left(1+\frac{q^2}{2m^2}\right)J_q,\nn
\TERMA^i_q&=&-\frac{4\alpha m^2T}{|\v{q}|}\left(1+\frac{q^2}{2m^2}\right)K_q,\nn
\TERMB^+_q&=&-\frac{4\alpha m^2T}{\pi|\v{q}|}\left(1+\frac{q^2}{4m^2}+\frac{q^0}{m}\right)I_q-\frac{2\alpha T^2}{\pi}\tilde I_q,\nn
\TERMB^-_q&=&-\frac{4\alpha m^2T}{|\v{q}|}\left(1+\frac{q^2}{4m^2}+\frac{q^0}{m}\right)J_q,\nn
\TERMB^i_q&=&-\frac{4\alpha m^2T}{|\v{q}|}\left(1+\frac{q^2}{4m^2}+\frac{q^0}{m}\right)K_q, 
\eea
where
\bea
I_q&=&\int_0^\infty\frac{dzz}{\tilde\omega_z}n_z\ln\left|\frac{r_z+z}{r_z-z}\right|,\nn
\tilde I_q&=&\int\frac{dzz^2}{\tilde\omega_z}n_z,\nn
J_q&=&\int_0^\infty\frac{dzz}{\tilde\omega_z}n_z\Theta(1-r_z)\mr{sign}(\mr{sign}(q^0)\tilde\omega_z+\beta q^0),\nn
K_q&=&-\int_0^\infty\frac{dzz}{\tilde\omega_z}\frac{n_z\Theta(1-r)}
{1+e^{-\sqrt{\tilde\omega_z^2+z^2+(\beta q^0)^2+2\beta q^0\tilde\omega_z}}},
\eea
with $r_z=\frac{q^2}{2|\v{q}|T}+\frac{q^0}{|\v{q}|}\tilde\omega_z$, $\tilde\omega_z=\sqrt{(\beta m)^2+z^2}$
and $n_z=1/(e^{\sqrt{(\beta m)^2+z^2}}+1)$. The results for finite density and vanishing temperature are
\bea\label{reimxm}
\TERMA^+_q&=&\frac{2\alpha k_F^2m}{\pi|\v{q}|}\left(1+\frac{q^2}{2m^2}\right)L_q,\nn
\TERMA^-_q&=&-\frac{\alpha k_F^2m}{|\v{q}|}\left(1+\frac{q^2}{2m^2}\right)M_q,\nn
\TERMB^+_q&=&\frac{2\alpha k_F^2m}{\pi|\v{q}|}\left(1+\frac{q^2}{4m^2}+\frac{q^0}{m}\right)L_q,\nn
\TERMB^-_q&=&-\frac{\alpha k_F^2m}{|\v{q}|}\left(1+\frac{q^2}{4m^2}+\frac{q^0}{m}\right)M_q,\nn
\TERMA^i_q&=&-\frac{\alpha k_F^2m}{|\v{q}|}\left(1+\frac{q^2}{2m^2}\right)N_q,\nn
\TERMB^i_q&=&-\frac{\alpha k_F^2m}{|\v{q}|}\left(1+\frac{q^2}{4m^2}+\frac{q^0}{m}\right)N_q
\eea
with
\bea\label{reimxf}
L_q&=&r+\hf(1-r^2)\ln\left|\frac{r+1}{r-1}\right|,\nn
M_q&=&\Theta(1-|r|)(1-r^2),\nn
N_q&=&\begin{cases}1-r^2&|\v{q}|>2k_F,~-1<r<1\cr1-r^2&|\v{q}|<2k_F,~-1-\frac{|\v{q}|}{k_F}<r<1\cr
\frac{(\omega+2m)\omega}{k_F^2}&|\v{q}|<2k_F,~-\frac{|\v{q}|}{2k_F}<r<1-\frac{|\v{q}|}{k_F}\end{cases}.
\eea
The variable $r$ in these equations is given by Eq. \eq{rdef} after the replacement
$k_{ch}=k_F$ and $\omega_p=m$.

$L$ is the relativistic generalization of the Lindhardt function. The last two 
expressions in Eqs. \eq{reimxm} give the relativistic version of the decoherence factor 
of the Coulomb field, considered in Ref. \cite{coulomb}. The expressions
\eq{reimxt}-\eq{reimxf} are valid for nonrelativistic electron gas but naturally
comprise the relativistic kinematics for photons.

\section{Reduced density matrix}\label{densmatr}
We use now the functional in Eq. \eq{wpint} to obtain the reduced density matrix
for the EM field. A single charged field is considered for simplicity and it will be
treated by CTP boundary conditions and the OTP boundary conditions are used for the EM field. 
The generator functional in Eq. \eq{wpint} with $\ha=\hj=0$, considered as a 
functional of the final field configurations gives the reduced density matrix
\be\label{rdenm}
\rho[A^+_{f~\v{x}},A^-_{f~\v{x}}]
=\int D[\hA]\biggr(\prod_n D[\hat\psi_n]D[\hat\psib_n]\biggl)
e^{i\sum_n\hat\psib_n[\hG^{-1}_{0n}-e_n\hsigma\hA\bre]\hat\psi_n
+\frac{i}{2}\hA\hD^{-1}_0\hA}.
\ee
The integration over the charged field can be carried out with the result
\be
\rho[A^+_{f~\v{x}},A^-_{f~\v{x}}]=\int D[\hat{A}]e^{iS_{eff}[\hA]}
\ee
where the effective bare action is
\be\label{otpact}
S_{eff}[\hA]=\hf\hA(\hD^{-1}_0-\hat\Pi)\hA+\ord{\hA^3},
\ee
$\hat\Pi$ being given by Eq. \eq{totse}. This functional integral contains 
information about the dynamics described by the density matrix in a manner similar 
the path integral describes the transition amplitude between pure states in the single 
time formalism. The real part $\Re S[\hA]$ is identical to $\Gamma[\hA]$, obtained 
as the Legendre transform of $\Re W[\ha,\hj]$ for $\ha=0$, it determines the 
expectation value of the EM field and will be used to extract the polarization 
induced by the charges. The imaginary part controls the width of the peak in the 
reduced density matrix, and the decoherence and consistency of the EM field.

The normal modes of the quadratic part of the action in the path integral
expression for the transition amplitudes are plane waves. It is easy to see
that $\hat\Pi_q$ is not diagonalizable in the time axis index only. The normal 
modes of the CTP action \eq{otpact} couple the wave vector $q=(\omega,\v{q})$ with the 
CTP index and are labeled by the wave vector, space-time and CTP indices. The dynamical 
role of each normal mode is characterized by three numbers,
\bea
\Pi^{n\mu\nu}_q&=&-\int dxe^{iqx}\Re i\la T[j^\mu_xj^\nu_0]\ra,\nn
\Pi^{f\mu\nu}_q&=&-\int dxe^{iqx}\Re i\la j^\mu_xj^\nu_0\ra,\nn
\Pi^{i\mu\nu}_q&=&\int dxe^{iqx}\Im i\la T[j^\mu_xj^\nu_0]\ra,
\eea
by applying the parametrization \eq{scnf} for the current-current two-point 
function $\htG$. It is easy to check that the Fourier transforms $\Pi^n$ and $\Pi^i$ 
are real and $\Pi^f$ is purely imaginary. In the one-loop approximation the real 
quantities arise from the interactions within a single time axis and the purely
imaginary quantity corresponds to correlation between the time axes. This
correlation, realized by Green functions connecting the two time axes,
arises because the charged particles which are exchanged in the process represented
by the Green functions are entangled with the EM field. 

We shall consider these quantities in the $(\v{q},\omega)$ space. The zeros of 
$\Pi^n_{\omega,\v{q}}$ locate the mass-shell of 
the quasiparticles, represented by the normal modes of the diagonal, $++$ or
$--$ blocks of the quadratic bare action. The imaginary part of the two-point
function in the space-time, which is given by $i\Pi^i_{\omega,\v{q}}$ in
Fourier-space, controls two, superficially different dynamical processes. 
On the one hand, the decay of the  quasiparticles, the inverse life-time of a 
quasiparticle can be identified by $\Pi^i_{\omega,\v{q}}$ evaluated on the 
mass-shell. On the other hand, according to the structure displayed in 
Eq. \eq{scnf}, this parameterizes the suppression of the off-diagonal matrix 
elements of the reduced density matrix, realized by the mode $\omega,\v{q}$ of 
the path integral. Thus the consistency and decoherence of the quasiparticle modes 
on the mass-shell have the same dynamical origin as the finite life-time of the 
quasiparticles. This is the expected relation between the dynamical breakdown 
of the time reversal invariance and the classical limit. 

We now turn to the discussion of the qualitative features of $\Pi^n$ and $\Pi^f$ 
in describing the polarizability of the charges and the impact of $\Pi^i$ on the 
classical limit.

\subsection{Electric and magnetic susceptibilities}
To make contact with the usual three-dimensional notation we introduce the parametrization
$A^\mu=(\phi,\v{A})$ of the vector potential, giving rise to 
$\v{E}=-\v{\nabla}\phi-\partial_0\v{A}$ and $\v{B}=\v{\nabla}\times\v{A}$. 
We consider quadratic action in the EM field in a formal manner, as a
book-keeping device for the equation of motion \eq{maxallp}, 
namely we assume that the actions contain the inverse of the retarded Green function.
On the one hand, we can introduce the retarded electric and magnetic 
susceptibilities $\chi$ and $\tilde\chi$, respectively with the phenomenological parametrization
\be\label{phenmaxw}
S_0[A]=\hf\int_{x,y}(\v{E}_x\epsilon_{xy}\v{E}_y-\v{B}_x\tilde\mu_{xy}\v{B}_y)
\ee
of the Maxwell action. On the other hand, the effective action for the EM field expectation 
value yielding the equation of motion is
\be\label{micrmaxw}
\Gamma=\hf AD^{r-1}A
\ee
where $D^{r-1}=D_0^{r-1}-e(\tG^{++}-\tG^{+-})e$ according to Eq. \eq{dbr}. Therefore the 
equivalence of the two quadratic form, expressed by the equation 
\be\label{dreqsfd}
\v{q}^2\begin{pmatrix}1&-\v{n}\nu\cr-\v{n}\nu&\nu^2\openone-\v{T}\end{pmatrix}
-\TERMB^r\begin{pmatrix}1&\v{n}\nu\cr\v{n}\nu&\nu^2L\end{pmatrix}
+\hf[\TERMB^r(1-\nu^2)-\TERMA^r]\begin{pmatrix}0&0\cr0&T\end{pmatrix}
=\v{q}^2\left[\epsilon\begin{pmatrix}1&-\v{n}\nu\cr-\v{n}\nu&\nu^2\openone-\v{T}\end{pmatrix}
-(\tilde\mu-\epsilon)\begin{pmatrix}0&0\cr0&\v{T}\end{pmatrix}\right],
\ee
follows in Fourier space with $X^r=X^{++}-X^{+-}$. The left hand side contains 
the free inverse propagator and the retarded self energy as given by Eqs. \eq{phse} 
and \eq{vachatab}-\eq{reimpa}. The right-hand side is $\delta^2S_0[A]/\delta A_{-q}\delta A_q$ 
with $\nu=\omega/|\v{q}|$, $\v{n}=\v{q}/|\v{q}|$, $\v{L}=\v{n}\otimes\v{n}$, and $\v{T}=\openone-\v{L}$.
We insert into the right-hand side the expressions $\epsilon=1+\chi$ and $\tilde\mu=1+\tilde\chi$
which gives, after canceling the free inverse propagator on the left-hand side by the
contributions of the one on the right-hand side
\bea\label{envssusc}
\chi_q&=&-\frac{\TERMB^r_q}{\v{q}^2},\nn
\tilde\chi_q&=&\frac{\TERMB^r_q(1-3\nu^2)-\TERMA^r_q}{2\v{q}^2}.
\eea

\begin{figure}
\parbox{8cm}{\includegraphics[scale=.7]{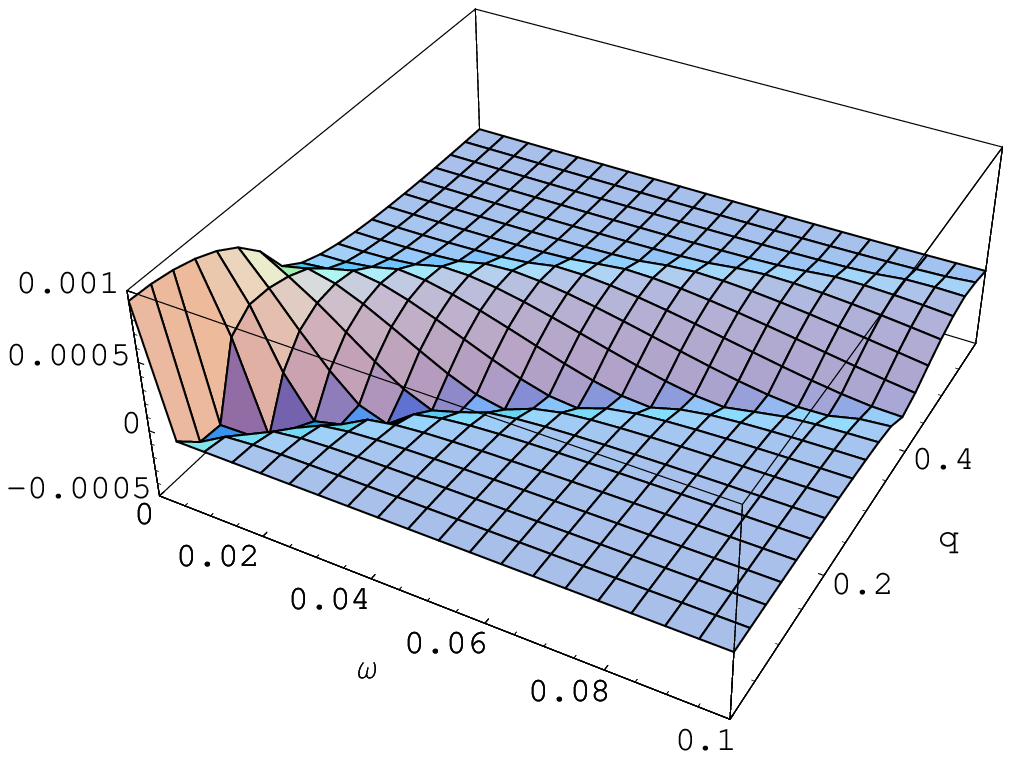}\\(a)}
\parbox{8cm}{\includegraphics[scale=.7]{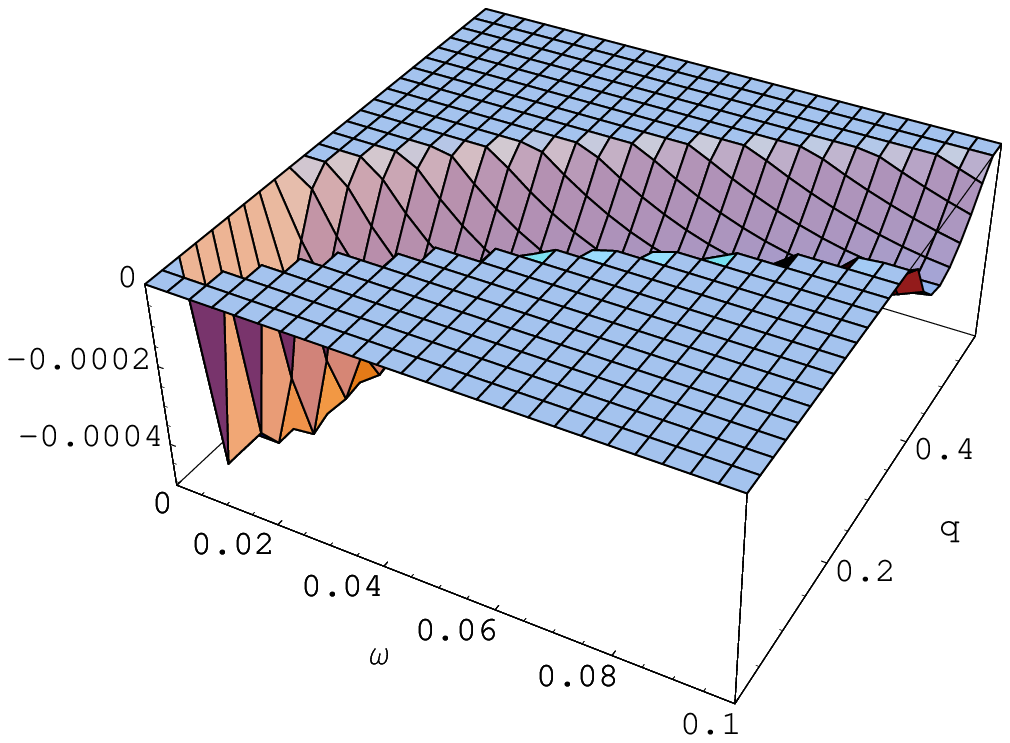}\\(b)}
\parbox{8cm}{\includegraphics[scale=.7]{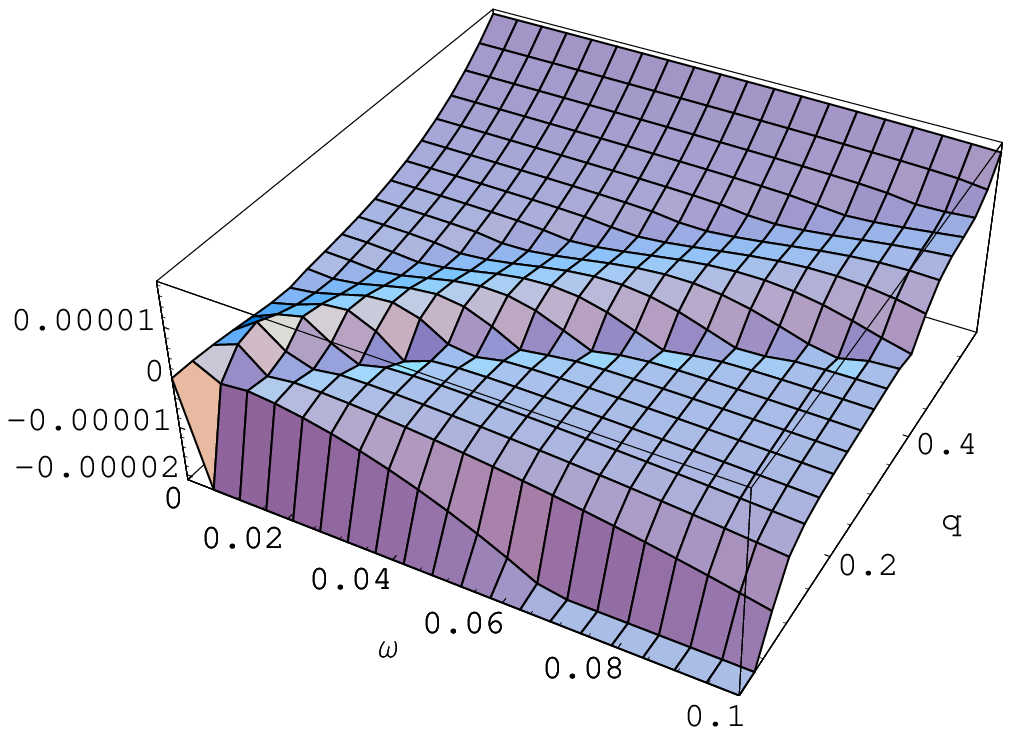}\\(c)}
\parbox{8cm}{\includegraphics[scale=.7]{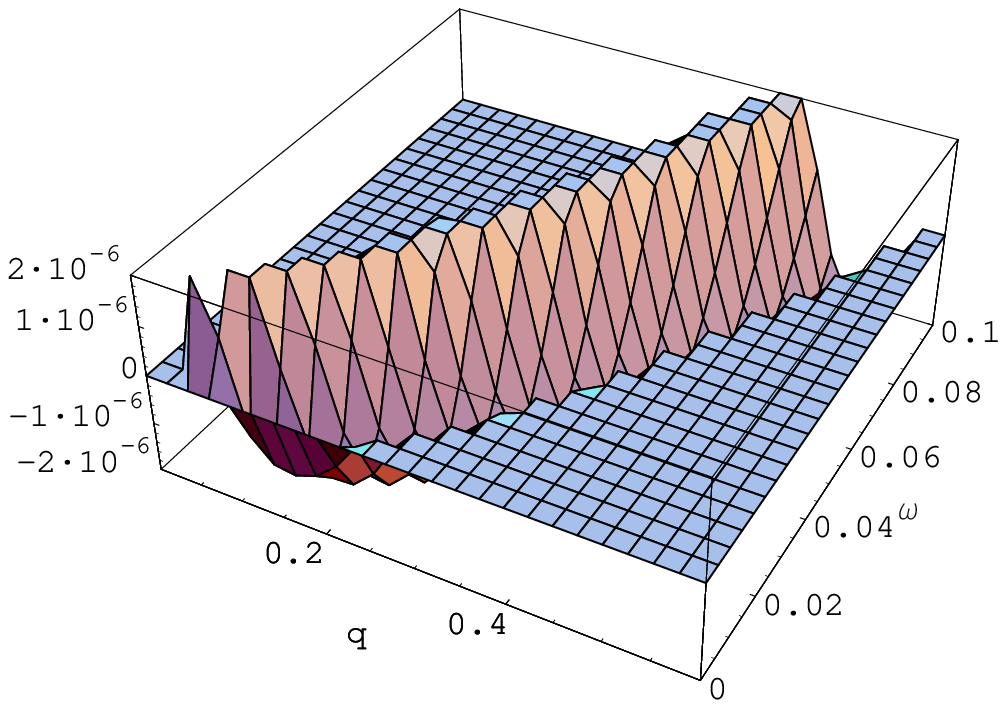}\\(d)}
\caption{The susceptibilities, (a):$\v{q}^2\Re\chi=\v{q}^2(\Re\epsilon-1)$, (b):$\v{q}^2\Im\epsilon$,
(c):$\v{q}^2\Re\tilde\chi=\v{q}^2(\Re\tilde\mu-1)$, (d):$\v{q}^2\Im\tilde\mu$ as functions of $\omega$ and 
$|\v{q}|$.}\label{susc}
\end{figure}

\begin{figure}
\parbox{5cm}{\includegraphics[scale=.45]{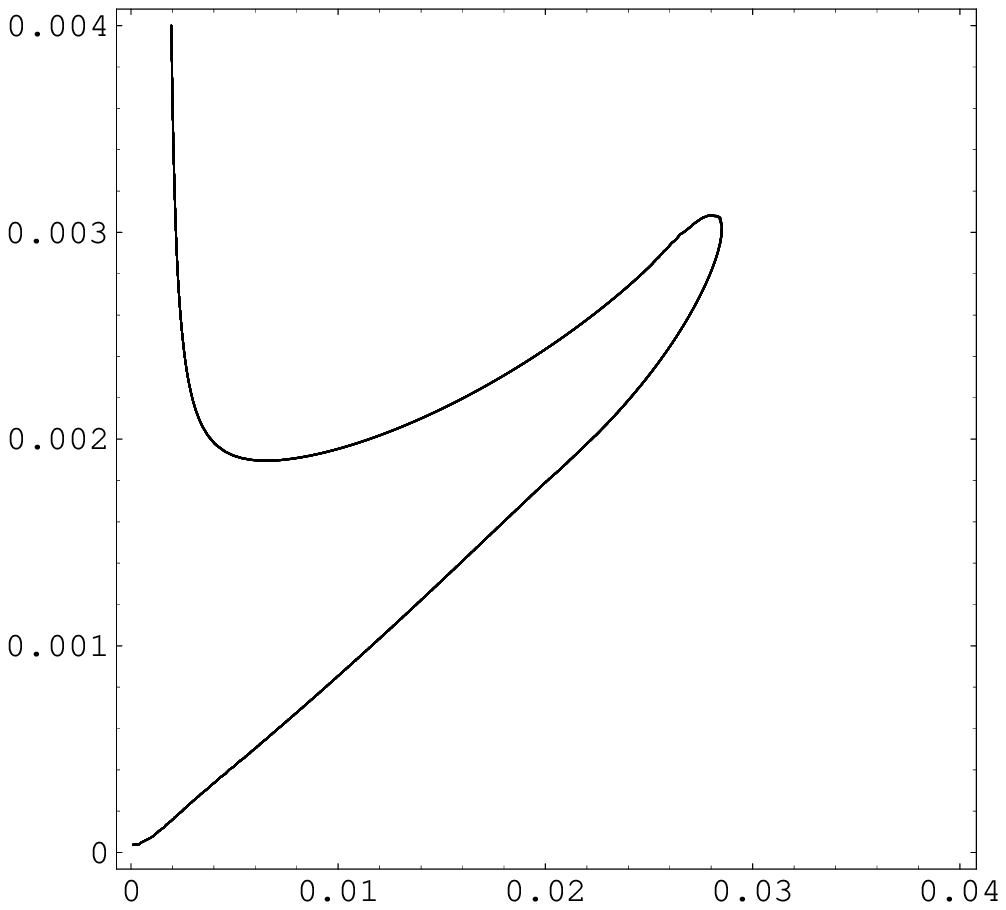}\\(a)}
\parbox{5cm}{\includegraphics[scale=.45]{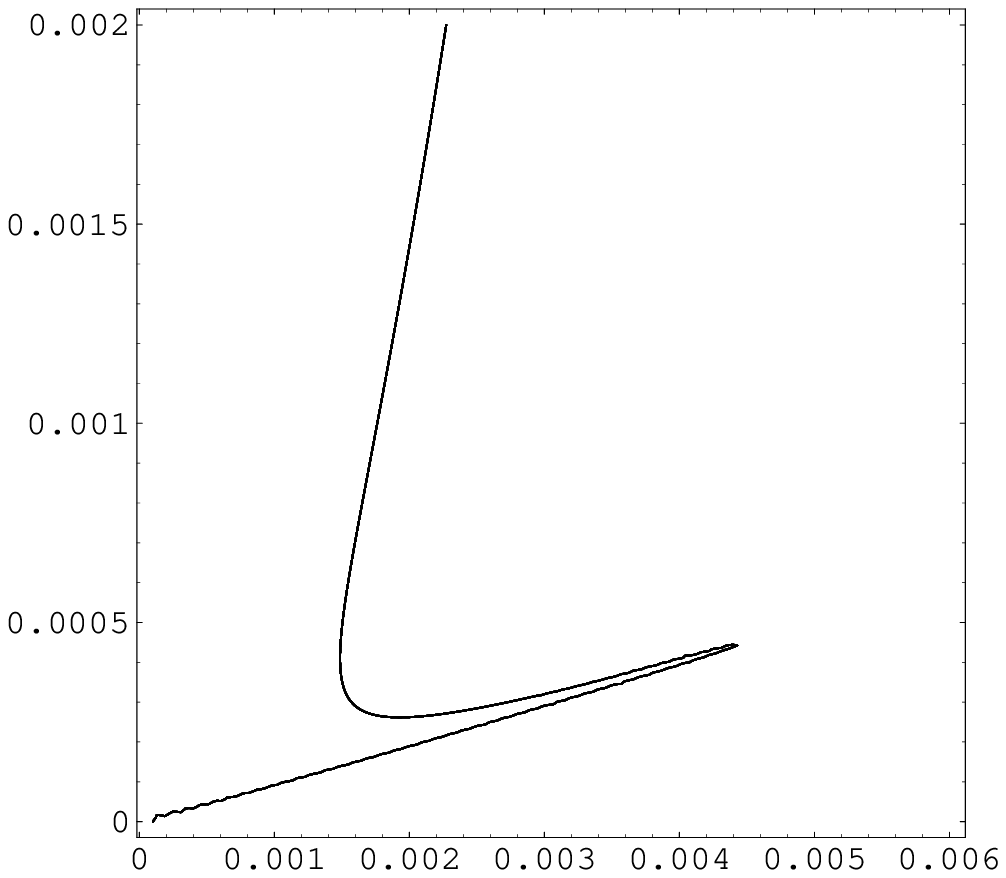}\\(b)}
\parbox{5cm}{\includegraphics[scale=.45]{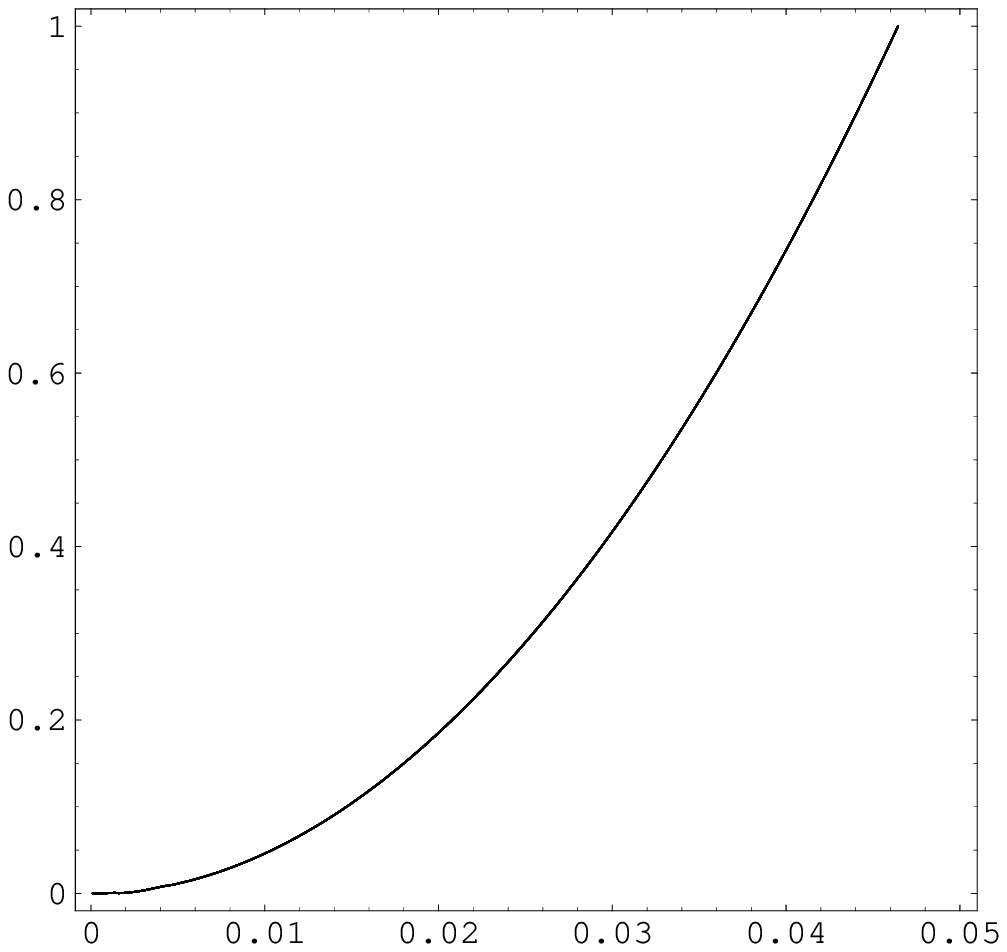}\\(c)}
\caption{Quasi-particles on the $(|\v{q}|,\omega)$ plane. 
(a): In the electric sector they are the roots of the equations $\Re\epsilon=0$. 
The zero sound and the plasmon modes correspond to the straight line
starting at the origin and after the turning point, respectively. 
The magnetic quasiparticles are at $\Re\tilde\mu=0$, (b) is around the origin and
(c) is on a larger part of the $(|\v{q}|,\omega)$ plane.}\label{contplm}
\end{figure}

Let us now see the more detailed, numerical features of the polarization functions
at finite density, obtained by using Eqs. \eq{reimx}, \eq{reimxm}-\eq{reimxf}
to compute $X^r_q=X^{++}_q-X^{+-}_q$ in Eqs. \eq{envssusc}. The susceptibilities display
a divergent structure which is well known for nonrelativistic Coulomb gas. The magnetic sector and the 
relativistic effects for the EM field are retained in this calculation which 
slightly modifies the picture. The susceptibilities, $\v{q}^2\epsilon_{\omega,\v{q}}$ and 
$\v{q}^2\tilde\mu_{\omega,\v{q}}$ are shown in Figs. \ref{susc} for $m=1$ and $\mu=0.1$.
They all take appreciable values for $|r|\approx1$, cf. Eqs. 
\eq{reimxm}-\eq{reimxf} and \eq{disprel}. The quasiparticle conditions 
$\Re\epsilon_{\omega,\v{q}}=0$, $\Re\tilde\mu_{\omega,\v{q}}=0$ have solution
along the valley of the line $r=1$, shown on the $(|\v{q}|,\omega)$ plane in
Fig. \ref{contplm}. The approximately linear parts in the electric and magnetic sectors
belong to the zero sound, the back turning sections correspond to the plasmonlike
excitations. The EM field may have relativistic energy and momenta and the relativistic
correction factors $\ord{q^2/2m^2}$ in Eqs. \eq{reimxm} modify the plasmon
lines in an essential manner. They prevent the plasmons from becoming long range and 
push the frequency up sharply as the wave vector tends to zero in the electric
sector, as seen in Fig. \ref{contplm} (a). For $\omega\ge2$ we naturally 
run into the pair creation singularities. The normal modes of the magnetic sector 
behave in a rather peculiar manner, the plasmon line, shown in Fig. \ref{contplm} (c) 
follows approximately a nonrelativistic dispersion relation curve of mass $m_{eff}=10^{-3}$
even in the relativistic domain.

\begin{figure}
\parbox{8cm}{\includegraphics[scale=.7]{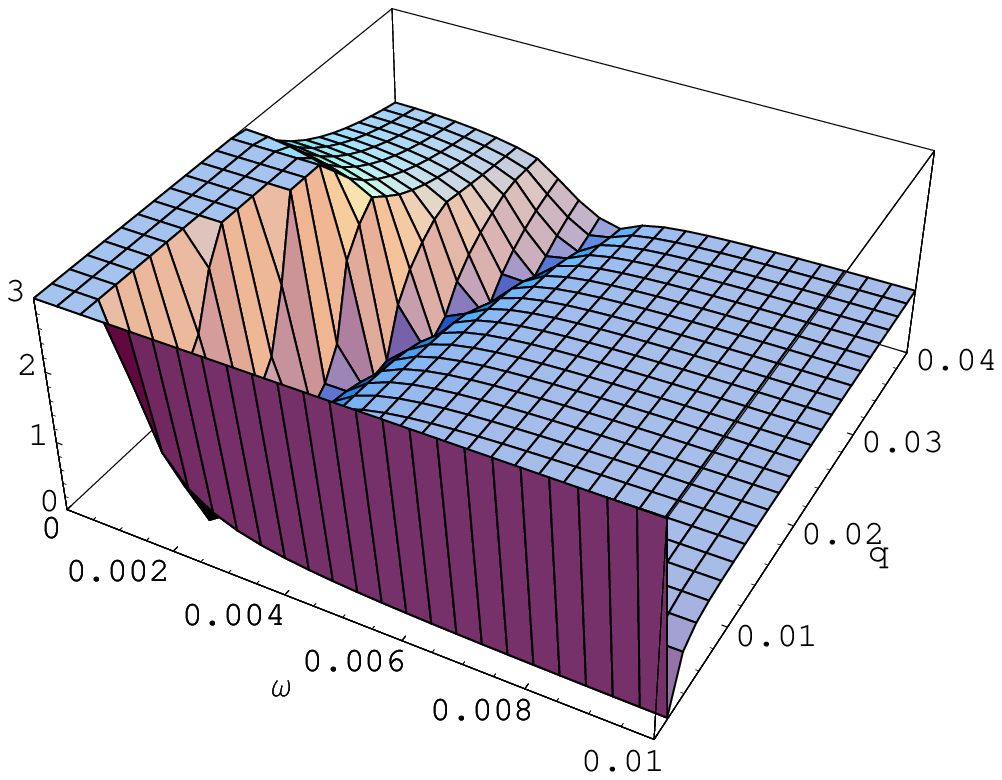}\\(a)}
\parbox{8cm}{\includegraphics[scale=.7]{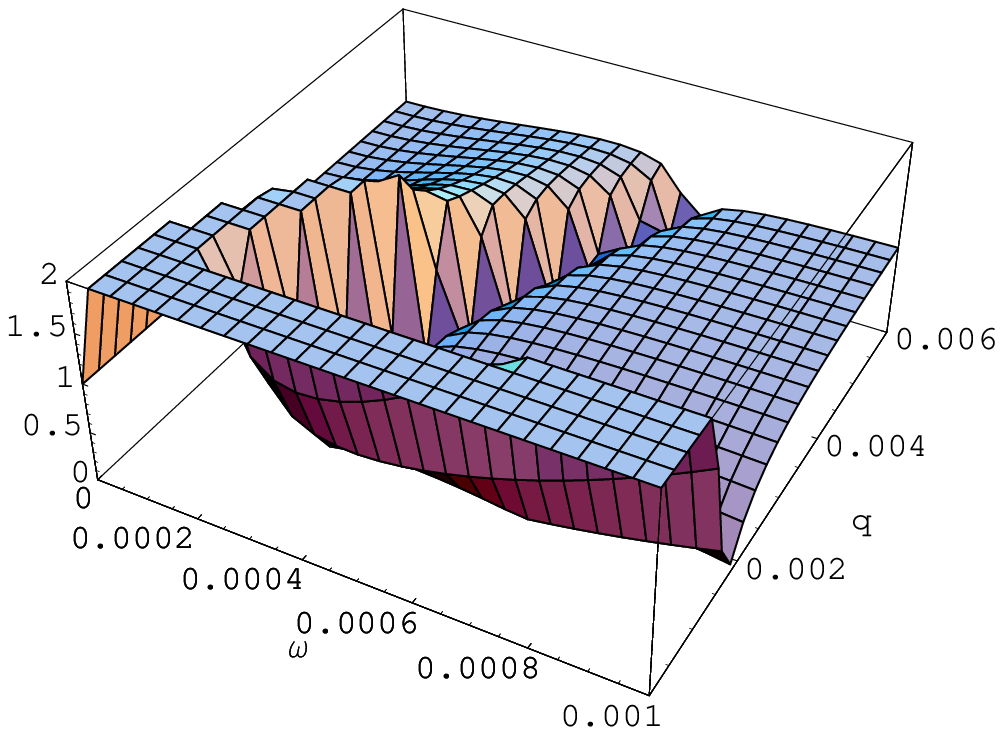}\\(b)}
\caption{Absolute magnitude of the polarizabilities. (a):The electric polarizability,
$|\epsilon|$, develops divergences as $\v{q}\to0$ and displays a finite peak for
$\omega\to0$ and $|\v{q}|<0.4$.
(b):$|\tilde\mu|$.}\label{absepsm}
\end{figure}

The screening of the electric sector manifests itself in the nonvanishing
value of the product $\v{q}^2\Re\epsilon$ in the limit $\v{q}\to0$ and the
resulting infrared divergence in $|\epsilon|$ as shown in Fig. \ref{absepsm} (a).
The valley in $|\epsilon|$ identifies the strongly coupled modes, they are
along the plasmon line of Fig. \ref{contplm} (a) and on the continuation of the 
zero-sound line. The zero-sound line shown in Fig. \ref{contplm} (a) is
weakly coupled due to the short life-time. The magnetic sector is qualitatively similar,
$|\tilde\mu|$ plotted in Fig. \ref{absepsm} (b) indicates strongly coupled magnetic
plasmons and weakly coupled magnetic zero sound modes. The magnetic field is screened
as well as shown by the nonvanishing value of $\v{q}^2\Re\tilde\mu$ in Fig. \ref{susc} (c)
in the infrared, $\v{q}\to0$ limit.

The physics is simpler at low temperature and vanishing density due to the 
gap in the excitation spectrum for the charges. One finds qualitatively similar
frequency and wave vector dependence as displayed in the figures above except
that the overall scale is suppressed by $\exp(-m/T)$.

\subsection{Consistency and decoherence}
\begin{figure}
\parbox{8cm}{\includegraphics[scale=.7]{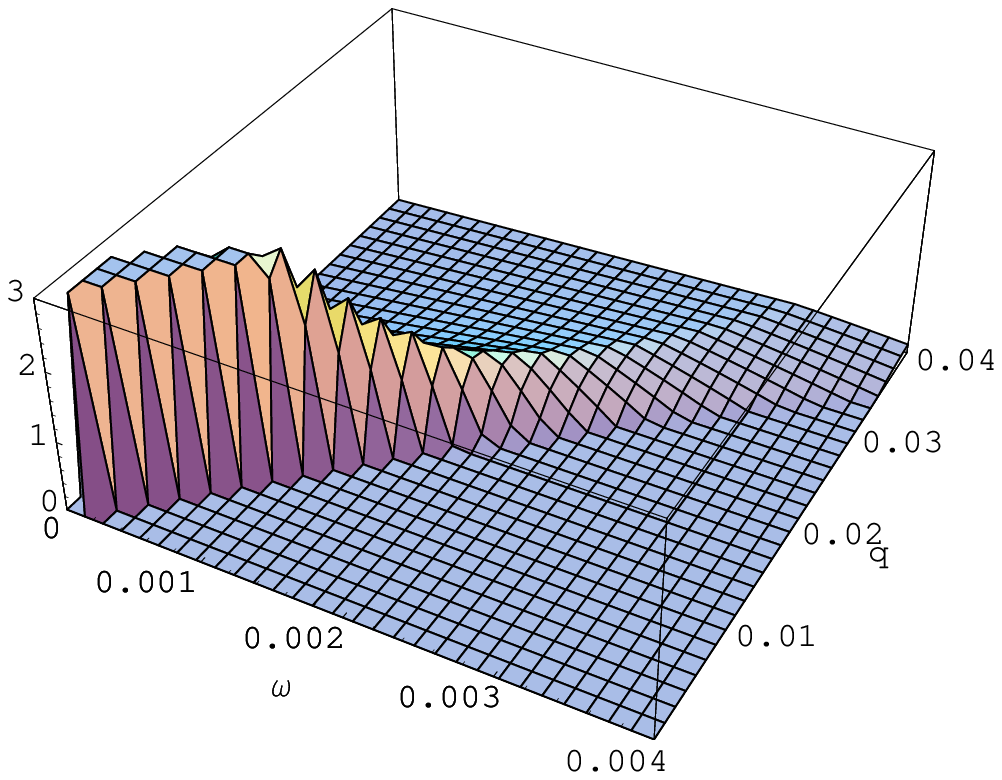}\\(c)}
\parbox{8cm}{\includegraphics[scale=.7]{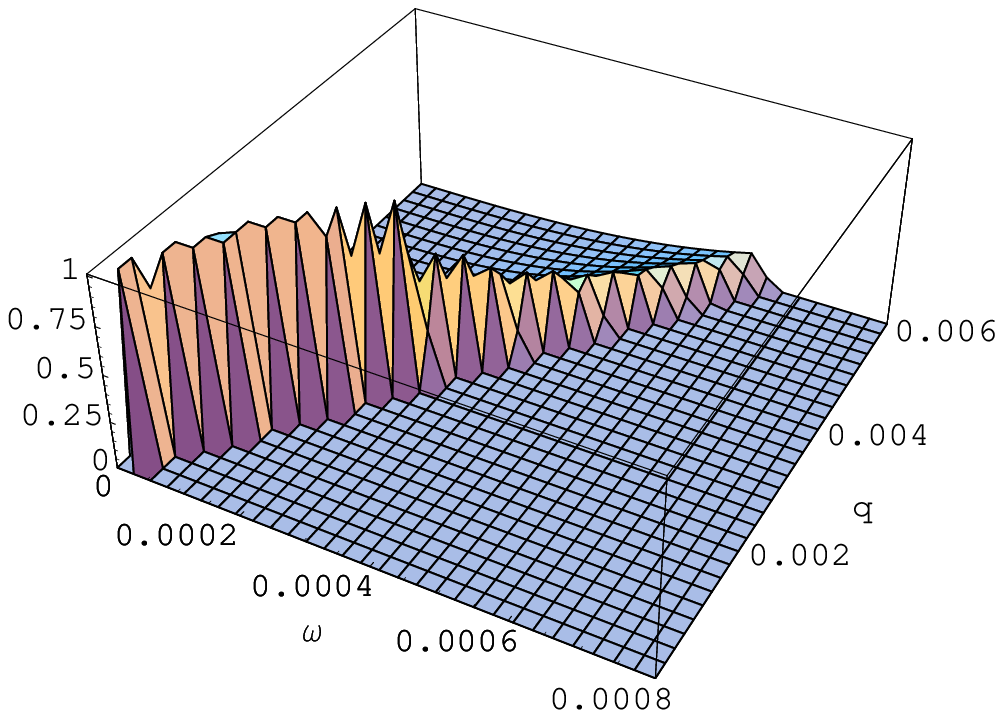}\\(d)}
\caption{The imaginary part of the bare effective action. (a):$\epsilon^d$, (b):$\tilde\mu^d$.}\label{dec}
\end{figure}

The reduced density matrix, given by Eq. \eq{rdenm} can always be diagonalized
and the resulting basis consists of completely decohered states. But one can
gain no more insight into the classical limit by using this basis. Instead, we are
interested in the role of physically motivated states which persist at
both side of the quantum-classical transition. These correspond to quasiparticles, 
defined by the $\ord{A^2}$ part of the bare action \eq{otpact}. Therefore, we seek
the contribution of the quasiparticles to decoherence in what follows.
It is easy to see by means of the structure \eq{scnf} of $\htG$ and the
expression \eq{totse} for the self energy that the imaginary part of the 
effective bare action $\Im S[\hA]$ of Eq. \eq{otpact}
depends on the combination $A^+-A^-$ of the EM field only. Eq. \eq{dreqsfd}
can be used again to parameterize it as
\be
\Im S[\hA]=\hf\int_x(\v{E}^d\epsilon^d\v{E}^d-\v{B}^d\tilde\mu^d\v{B}^d)
\ee
with $\v{E}^d=\v{E}^+-\v{E}^-$, $\v{B}^d=\v{B}^+-\v{B}^-$ where the decoherence parameters
\bea
\epsilon^d_q&=&-\frac{\TERMB^{i}_{q}}{\v{q}^2}\nn
\tilde\mu^d_q&=&\frac{\TERMB^{i}_{q}(1-3\nu^2)-\TERMA^{i}_{q}}{\v{q}^2},
\eea
are expressed in terms of the imaginary parts introduced in Eq. \eq{reimx} and are
plotted in Fig. \ref{dec}.
$\Im S[\hA]$ governs the suppression of each plane wave mode along the off-diagonal
direction in the reduced density matrix. It is easy to see that this functional is
semidefinitive because it measures the phase space available for particle-antiparticle 
or particle-hole excitations which incorporate decoherence. Its value, taken on the
quasiparticle line of Fig. \ref{contplm}, yields the inverse life-time.
We see that the zero-sound quasiparticles are rendered unimportant by their short 
life-time. Naturally, higher loop contributions
to the effective action generate imaginary part and finite life-time
beyond the zero-sound line by the multi particle-antiparticle or particle-hole
excitations, but this remains a weak effect.

When considered from the point of view of the quantum-classical
transition, the effective bare action monitors the impact of the entanglement between the
EM field and the charges. One measure of this entanglement is decoherence which
corresponds to an instantaneous state. But we find more information in this action.
The suppression of the contributions in the path integral representation of the 
density matrix as the functional of the difference of the two field trajectories 
carries information about the consistency of histories of the EM field \cite{grif}. 
The contribution of a plane wave, a normal mode of the one-loop dynamics of the 
EM field, to the conditional probability distributions is closer to the one expected in 
the standard, classical probability if $\Im S[\hA]$ increases with $A^+-A^-$ because 
the quantum interference term which violates the additivity of probabilities is 
suppressed by $e^{-\Im S}$ \cite{hig}. 

The lesson is that the classical probabilities are recovered by plane wave modes whose
dispersion relation is that of the short life-time quasiparticles. Their short decay time assures the dynamical breakdown of the time
reversal invariance, needed for the classical limit. These modes which maximize
$\Im S$ are located around the long wavelength part of the zero sound curve at finite 
density and vanishing temperature. The suppression of the off-diagonal elements of
the reduced density matrix for these plane wave EM fields shows that these 
field configurations become good pointer states in the classical limit.

\section{Conclusions}\label{concl}
The problem of deriving classical electrodynamics from QED is pursued further
in this work. It is based on the monitoring of subclassical fields, and the
expectation values of local operators with space-time resolution limited
formally by the UV cutoff of QED. The final goal is to weaken the resolution and
thereby recover the macroscopic, classical dynamics for the
space-time dependence of the expectation values. As a preparatory step,
we make a coarse graining by eliminating the charged degrees of freedom with a fixed, microscopic
space-time resolution. Polarization effects, arising in this manner,
reflect competing coherent and decoherent phenomena, both at the
elementary and at the collective level. 

The double-time axes formalism used in this work was motivated by the implementation 
of initial conditions, as opposed to the boundary value problems within the variational method.
This step leads to the double-time axis formulation in a natural manner. 
The double-time axes formalism is well suited to 
follow the  eventual breakdown of time reversal invariance at the quantum level, 
driven dynamically by degeneracies or externally, by the boundary conditions in time. 
The breakdown of the time reversal invariance appears at the level of the 
expectation value of the EM field as the radiation field. A simple microscopic
way of separating the EM field into near and far components emerges as well
for they represent pure state self-interactions for the charges and 
generate entanglement between charges and the EM field, respectively.

Some well-known issues appear in a new point of view, 
such as the origin of the dynamics for the free-field components in 
the variational equations. The free-field component belongs to the null space of 
the kernel of the free action and drops out from the linearized equation of motion.
This problem is avoided in the double-time formalism, and Feynman's $i\epsilon$ 
prescription seems to be the analogous treatment in the single time formalisms of 
both quantum and classical field theory.

Another question which can easily be addressed is the relation between
the classical limit and the dynamical breakdown of the time reversal invariance.
The block structure of the CTP propagators shows that a shorter lifetime of the 
quasiparticles means more consistent EM field trajectories and a narrower peak
in the absolute magnitude of the reduced density matrix $|\rho(x,y)|$
around the diagonal, $x=y$. Another way to see this relation by means of the EM 
field expectation value is to note that the purely imaginary far field component 
of the CTP polarization tensor, $\Pi^f_{\omega,\v{q}}$, plays a double role in the 
dynamics. On the one hand, it characterizes the mixed 
states arising from the charge-EM field entanglement, and on the other hand, it leads 
to a phase delay in the propagation of the EM field and thereby breaks the time 
reversal invariance.

These results open the way for further developments, and we close by mentioning
some of them. The subclassical fields are expectation values but they are not
necessarily classical quantities because they may display the microscopic structure
in space-time where decoherence is not efficient. In order to identify initial conditions
which lead to classical fields, we need to control either the decoherence itself or
the characteristic scale of the field expectation value. The latter is determined
by the two-point functions and its handling requires the extension of the present scheme to 
external sources coupled to the bi-local product of the elementary fields. Such
an explicit treatment of the correlation functions would, in addition, allows us to 
test different collapse scenarios of the wave function.

Another extension of the present study is the detailed analysis of the effective action 
for the EM field and the electric current to derive a variational scheme for 
magnetohydrodynamics. The more accurate determination of the susceptibilities from first 
principles may lead to new experimental devices to identify some signatures of the 
underlying microscopic quantum theory by precision measurements in classical electrodynamics.

\acknowledgments
We thank J\'anos Hajdu for several stimulating discussions and Mahmud Merad for
his help in the initial phase of this project.

\appendix
\section{CTP propagators in the vacuum}\label{ctppropa}
The CTP vacuum propagators for scalar, vector, and spin-half particles are summarized in this Appendix.
The relation between expectation values and the parametrization of the propagator is achieved
by path integration, and the explicit expression, for the different expectation
values are computed in the operator formalism.

\subsection{Neutral bosons}
We introduce the propagator for a boson field $\phi$ with or without vector indices by means of the free generator functional,
\bea\label{freephotctp}
e^{iW[\hj]}&=&\int D[\hat\phi]e^{\frac{i}{2}\hat\phi\cdot\hD^{-1}\cdot\hat\phi+i\hj\cdot\hat\phi}\nn
&=&e^{-\frac{i}{2}\hj\cdot\hD\cdot\hj}
\eea
which gives, in the Heisenberg representation,
\bea\label{propid}
i\fdd{W[\hj]}{ij^+_a}{ij^+_b}
&=&\sum_n\la0||n\ra\la n|T[\phi_a\phi_b]|0\ra
=\la0|T[\phi_a\phi_b]|0\ra=iD^{++}_{ab},\nn
i\fdd{W[\hj]}{ij^-_a}{ij^-_b}&=&\sum_n\la0|\bar T[\phi_a\phi_b]|n\ra\la n|0\ra
=\la0|T[\phi_b\phi_a]|0\ra^*=iD^{--}_{ab},\nn
i\fdd{W[\hj]}{ij^+_a}{ij^-_b}&=&\sum_n\la0|\phi_b|n\ra\la n|\phi_a|0\ra
=\la0|\phi_b\phi_a|0\ra=iD^{+-}_{ab},\nn
i\fdd{W[\hj]}{ij^-_a}{ij^+_b}&=&\sum_n\la0|\phi_a|n\ra\la n|\phi_b|0\ra
=\la 0|\phi_a\phi_b|0\ra=iD^{-+}_{ab},
\eea
written as
\be\label{scctpprop}
i\begin{pmatrix}D&D^{+-}\cr D^{-+}&D^{--}\end{pmatrix}_{ab}
=i\begin{pmatrix}D&-D^{-+*}\cr D^{-+}&-D^\dagger\end{pmatrix}_{ab}
=\begin{pmatrix}\la0|T[\phi_a\phi_b]|0\ra&\la0|\phi_b\phi_a0|0\ra\cr
\la0|\phi_a\phi_b|0\ra&\la0|T[\phi_b\phi_a]|0\ra^*\end{pmatrix}
\ee
by means of the notation $D^{++}=D$. The basic CTP identity,
\be
T[\phi_a\phi_b]+\bar T[\phi_a\phi_b]=\phi_a\phi_b+\phi_b\phi_a,
\ee
written as
\be\label{ctpidhf}
D-D^\dagger=D^{-+}-D^{-+*}
\ee
allows us to parameterize the propagator in terms of three real functions, the near and
far field Green functions, $D^n$ and $D^f$, respectively, and the common imaginary part, $D_i=\Im D$,
as in Eq. \eq{scnf}. The detailed form of the real and imaginary parts,
\bea\label{vkpmdef}
\Re D_{aa'}&=&-\hf\mr{sign}(t-t')i\la0|[\phi_a,\phi_{a'}]|0\ra=D^n_{aa'},\nn
D_{iaa'}&=&-\hf\la0|\{\phi_a,\phi_{a'}\}|0\ra,\nn
\Re D^{-+}_{aa'}&=&-\frac{i}2\la0|[\phi_a,\phi_{a'}]|0\ra=D^f_{aa'},
\eea
justifies the introduction of retarded and advanced propagators
\bea\label{ragrenfct}
D^{\stackrel{r}{a}}_{aa'}&=&D^n_{aa'}\pm D^f_{aa'}\nn
&=&\mp\Theta(\pm(t-t'))i\la0|[\phi_a,\phi_{a'}]|0\ra.
\eea

The expectation values of the propagator \eq{scctpprop} are easiest to compute in the operator
formalism by means of the quantum field
\be
\phi(x)=\int_{\tilde{\v{k}}}[a(\v{k})e^{-ikx}+a^\dagger(\v{k})e^{ikx}],
\ee
where the notation
\be
\int_{\tilde{\v{k}}}=\int\frac{d^3k}{(2\pi)^32\omega_\v{k}}
\ee
$\omega_\v{k}=\sqrt{m^2+\v{k}^2}$ has been introduced. The nonvanishing canonical commutation relation for the creation and 
annihilation operators,
\be\label{canconrnb}
[a(\v{p}),a^\dagger(\v{p}')]=(2\pi)^32\omega_\v{p}\delta(\v{p}-\v{p}'),
\ee
gives, in a trivial manner,
\be\label{hsdpm}
iD^{\pm\mp}(x,x'))=\int_{\tilde{\v{k}}}e^{\pm ik(x-x')}
\ee
for a scalar particle. The multiplication by the Heaviside function,
\bea
\Theta(t-t')iD^{-+}(x,x'))&=&-\frac1i\int_\omega\frac{e^{-i\omega(t-t')}}{\omega+i\epsilon}
\int_{\tilde{\v{k}}}e^{-i\omega_\v{k}(t-t')+i\v{k}(\v{x}-\v{x'})}\nn
&=&i\int_{\tilde{\v{k}},\omega}\frac{e^{-i\omega(t-t')+i\v{k}(\v{x}-\v{x'})}}{\omega-\omega_\v{k}+i\epsilon},\nn
\Theta(t'-t)iD^{+-}(x,x'))&=&-\frac1i\int_\omega\frac{e^{-i\omega(t'-t)}}{\omega+i\epsilon}
\int_{\tilde{\v{k}}}e^{-i\omega_\v{k}(t'-t)+i\v{k}(\v{x}'-\v{x})}\nn
&=&-i\int_{\tilde{\v{k}},\omega}\frac{e^{-i\omega(t-t')+i\v{k}(\v{x}-\v{x'})}}
{\omega+\omega_\v{k}-i\epsilon}
\eea
gives the usual causal propagator and, we finally find
\be\label{scctpprope}
\begin{pmatrix}D&D^{+-}\cr D^{-+}&D^{--}\end{pmatrix}_k
=\begin{pmatrix}\frac{1}{k^2-m^2+i\epsilon}&-2\pi i\delta(k^2-m^2)\Theta(-k^0)\cr
-2\pi i\delta(k^2-m^2)\Theta(k^0)&-\frac{1}{k^2-m^2-i\epsilon}\end{pmatrix}
\ee
at vanishing temperature.

The Fourier transforms of the real and imaginary parts of the causal propagator are
\bea\label{ftrsprop}
D^n_k&=&P\frac{1}{k^2-m^2},\nn
D_{ik}&=&-\pi\delta(k^2-m^2).
\eea
In computing the real and imaginary parts of $D^{-}$, one has to keep in mind that they are
defined in Eqs. \eq{vkpmdef} in the real space-time as opposed to Fourier space, and the expression
\be
D^{-+}=-\pi i\int_ke^{-ik(x-x')}\delta(k^2-m^2)[\Theta(k^0)-\Theta(-k^0)]
-\pi i\int_ke^{-ik(x-x')}\delta(k^2-m^2)[\Theta(k^0)+\Theta(-k^0)]
\ee
yields
\be
D^f_k=-\pi i\delta(k^2-m^2)\epsilon(k^0)
\ee
and
\be
D^{\stackrel{r}{a}}_k=\frac{1}{k^2-m^2\pm i\epsilon\mr{sign}(k^0)}.
\ee

Note the remarkable feature of the near field propagator $D^n$ in the first expression of Eqs. \eq{ftrsprop}.
It is the only part of the CTP propagator which is nonvanishing off the mass shell, the support 
of the other functions $D_i$ and $D^f$ is just the mass shell. Furthermore, $D^n_k$
diverges in momentum space as we approach the mass shell, $k^2\to m^2$; therefore
\be\label{dnifn}
D^{n-1}D^f=0,
\ee
cf. Eq. (D11) in \cite{pol} for the regulated expression.

\subsection{Charged bosons}
In the case of a non-Hermitian boson field the free generator functional is
\be
e^{iW[\hj^\dagger,\hj]}=\int D[\hat\phi^\dagger]D[\hat\phi]e^{i\hat\phi^\dagger\cdot\hD^{-1}\cdot\hat\phi
+i\hj^\dagger\cdot\hat\phi+i\hat\phi^\dagger\cdot\hj}
=e^{-i\hj^\dagger\cdot\hD\cdot\hj}.
\ee
and the CTP propagator block-matrix elements, identified in a manner analogous to Eqs. \eq{propid}, are
\be
i\begin{pmatrix}D&D^{+-}\cr D^{-+}&D^{--}\end{pmatrix}
=\begin{pmatrix}\la0|T[\phi_a\phi^\dagger_b]|0\ra&\la0|\phi^\dagger_b\phi_a|0\ra\cr
\la0|\phi_a\phi^\dagger_b|0\ra&\la0|T[\phi_b\phi^\dagger_a]|0\ra^*\end{pmatrix}
\ee
and they can be written as in Eq. \eq{scnf} due to the CTP identity
\bea
T[\phi_x\phi^\dagger_{x'}]+T^*[\phi_x\phi^\dagger_{x'}]&=&\phi^\dagger_{x'}\phi_x+\phi_x\phi^\dagger_{x'}\nn
D-D^\dagger&=&D^{-+}+D^{+-}.
\eea

It is the charge conjugation invariance which guarantees that the detailed form of the 
functions $D_i$, $D^n$ and $D^f$ is the same as in the case of neutral particles. The
non-Hermitian quantum field 
\be
\phi(x)=\int_{\tilde{\v{k}}}[a(\v{k})e^{-ikx}+b^\dagger(\v{k})e^{ikx}]
\ee
involves the creation and annihilation operators whose nonvanishing canonical commutators are
\be\label{nhscommrel}
[a(\v{p}),a^\dagger(\v{p}')]=[b(\v{p}),b^\dagger(\v{p}')]=(2\pi)^32\omega_p.
\ee
These operators transform under the charge conjugation transformation $C$, $C^2=\openone$,
as $C^\dagger a(\v{p})C=b(\v{p})$, $C^\dagger b(\v{p})C=a(\v{p})$ and $C^\dagger \psi_xC=\psid_x$.
Now, the relation $D^{+-}=D^{-+}$ actually establishes the equivalence of vacuum expectation 
values computed in the particle and the antiparticle sectors, and it originates from the
charge conjugation invariance of the commutation relations \eq{nhscommrel} and of the 
vacuum. When the propagator is expressed in terms of $D$ and $D^{-+}$, the complex conjugation 
like in Eq. \eq{ctpidhf} reverses the order of $\psi$ and $\psid$, and one needs another
complex conjugation, this time embedded into the charge conjugation operator, to place the 
dagger at the appropriate operator in $D^{+-}$,
\be
iD^{+-}=\la0|\phi^\dagger_b\phi_a|0\ra=\la0|C^\dagger\phi_a\phi^\dagger_bC|0\ra^*=iD^{-+C*}.
\ee
In the presence of a noncharge conjugate invariant vacuum, $D^{+-}\not=D^{-+}$.

\subsection{Fermions}
The generator functional for the propagator,
\be
e^{iW[\hj,\bar{\hj}]}=\int D[\hat\psi]D[\hat\psib]
e^{i\hat\psib\cdot\hG^{-1}\cdot\hat\psi+i\bar{\hj}\cdot\hat\psi+i\hat\psib\cdot\hj}
=e^{-i\bar{\hj}\cdot\hG\cdot\hj},
\ee
leads to
\be
i\begin{pmatrix}G&G^{+-}\cr G^{-+}&G^{--}\end{pmatrix}^{\alpha\beta}_{xy}
=\begin{pmatrix}\la0|T[\psi^\alpha_x\psib^\beta_y]|0\ra&\la0|(\psib_y^\beta\psi_x^\alpha|0\ra\cr
-\la0|\psi^\alpha_x\psib^\beta_y|0\ra&\la0|T[(\gamma^0\psi_y)^\beta(\psib_x\gamma^0)^\alpha]|0\ra^*\end{pmatrix}
\ee
by following the strategy of Eqs. \eq{propid}. The CTP identity,
\be\label{ctpidferm}
T[\psi\psib]+\bar T[\psi\psib]=-\psib\psi+\psi\psib
\ee
now leads to the parametrization
\be
\begin{pmatrix}G&G^{+-}\cr G^{-+}&G^{--}\end{pmatrix}
=\begin{pmatrix}G^n+iG_i&G^f-iG_i\cr-G^f-iG_i&-G^n+iG_i\end{pmatrix}.
\ee

The field operator is written as
\be\label{fermfi}
\psi(x)=\int_\v{k}\frac{m}{\omega_k}\sum_\alpha[b_\alpha(\v{k})u^{(\alpha)}(\v{k})e^{-ikx}
+d^\dagger_\alpha(\v{k})v^{(\alpha)}(\v{k})e^{ikx}]
\ee
where $(k\br-m)u(k)=(k\br+m)v(k)=0$ and the creation and annihilation operators are
defined by their nonvanishing anti-commutation relations 
\be
\{b_\alpha(\v{p}),b^\dagger_\beta(\v{p}')\}=\{d_\alpha(\v{p}),d^\dagger_\beta(\v{p}')\}
=(2\pi)^3\frac{\omega_p}{m}\delta_{\alpha,\beta}\delta(\v{p}-\v{p}').
\ee
The charge conjugation is represented by $C^{-1}\psi C=U\psi^*$ with $U=i\gamma^2$
and the identity \eq{ctpidferm} allows us to write
\bea
i(G^{+-})^{\alpha\beta}_{xy}&=&(U\la0|C^\dagger\psi_x\psib_yC|0\ra^*U)^{\alpha\beta}
=(UiG^{-+C}U)^{*\alpha\beta}_{xy}\nn
iG^{--\alpha\beta}_{xy}&=&(\gamma^0iG_{yx}\gamma^0)^{*\beta\alpha}.
\eea

The Fourier decomposition \eq{fermfi} gives the explicit expression
\be
\begin{pmatrix}G&G^{+-}\cr G^{-+}&G^{--}\end{pmatrix}
=(i\partial\br_x+m)\begin{pmatrix}D&-D^{+-}\cr-D^{-+}&D^{--}\end{pmatrix},
\ee
where the scalar propagator in the right hand side is given by Eq. \eq{scctpprope}, in a standard manner.

\section{CTP propagators at finite temperature and density}\label{ctppropenv}
We extend the propagators recorded in the previous appendix for the presence
of heat and particle reservoirs.

\subsection{Neutral bosons}
It is advantageous to use a finite quantization box, where the commutation relation \eq{canconrnb} reads
\be
[a(\v{p}),a^\dagger(\v{p}')]=2V\omega_\v{p}\delta_{\v{p},\v{p}'}
\ee
and the quantum field becomes
\be
\phi(x)=\frac{1}{2V\omega_\v{k}}\sum_\v{k}[a(\v{k})e^{-ikx}+a^\dagger(\v{k})e^{ikx}].
\ee
The relation
\bea
\Tr[e^{-\beta H_0}\phi(x)\phi(x')]&=&\sum_{\{n\}}\sum_\v{k}\frac{e^{-ik(x-x')}}{2V\omega_\v{k}}\la\{n\}|e^{-\beta H}|\{n\}\ra\nn
&&+\sum_{\{n\}}\sum_\v{k}\frac{1}{4V^2\omega^2_\v{k}}(e^{ik(x-x')}+e^{-ik(x-x')})
\la\{n\}|e^{-\beta H}a^\dagger(\v{k})a(\v{k})|\{n\}\ra\nn
&=&\int_{\tilde{\v{k}}}e^{-ik(x-x')}\prod_\v{k}\sum_ne^{-\beta n\epsilon_\v{k}}
+\sum_{\{n\}}\sum_\v{k}\frac{1}{2V\omega_\v{k}}(e^{ik(x-x')}+e^{-ik(x-x')})\la\{n\}|e^{-\beta H}n(\v{k})|\{n\}\ra\nn
&=&\prod_\v{k}\frac{1}{1-e^{-\beta\epsilon_\v{k}}}\left[iD^{-+}_{T=0}(x,x')
+\int_{\tilde{\v{k}}}(e^{ik(x-x')}+e^{-ik(x-x')})\frac{\sum_nne^{-\beta\epsilon_\v{k}}}{\sum_ne^{-\beta\epsilon_\v{k}}}\right],
\eea
where $H_0$ is the free Hamiltonian gives
\bea
iD^{-+}(x,x')&=&\frac{\Tr[e^{-\beta H}\phi(x)\phi(x')]}{\Tr[e^{-\beta H}]}\nn
&=&iD^{-+}_{T=0}(x,x')+\Delta D(x,x'),
\eea
with $D^{-+}_{T=0}$ being taken from Eq. \eq{hsdpm} and
\be
\Delta D=\int_ke^{-ik(x-x')}\frac{2\pi\delta(k^2-m^2)}{e^{\beta\epsilon_\v{k}}-1}.
\ee
Similar steps produce
\bea
iD^{+-}(x,x')&=&\frac{\Tr[e^{-\beta H}\phi(x')\phi(x)]}{\Tr[e^{-\beta H}]}\nn
&=&iD^{+-}_{T=0}(x,x')+i\Delta D(x,x').
\eea
The lesson is that finite temperature modifies the imaginary part of the propagator only, where
\be
\Delta D_{ik}=-\frac{2\pi\delta(k^2-m^2)}{e^{\beta\epsilon_\v{k}}-1}
\ee
and $\Delta D^R=\Delta D^A=0$. The independence of the retarded and advanced propagators
from the temperature can be understood by recalling that these propagators correspond to
fixed initial or final conditions.

\subsection{Charged bosons}
The non-Hermitian field
\be
\phi(x)=\frac{1}{2V\omega_\v{k}}\sum_\v{k}[a(\v{k})e^{-ikx}+b^\dagger(\v{k})e^{ikx}]
\ee
involving the operators which satisfy the nonvanishing commutation relations
\be
[a(\v{p}),a^\dagger(\v{p}')]=[b(\v{p}),b^\dagger(\v{p}')]=2V\omega_\v{p}\delta_{\v{p},\v{p}'}
\ee
give rise to 
\bea\label{mpcb}
\Tr[e^{-\beta(H-\mu Q)}\phi(x)\phi^\dagger(x')]&=&\int_{\tilde{\v{k}}}e^{-ik(x-x')}\prod_\v{k}\sum_ne^{-\beta n(\epsilon_\v{k}-\mu)}\\
&&+\sum_{\{n\}}\sum_\v{k}\frac{1}{2V\omega_\v{k}}\la\{n\}|e^{-\beta(H-\mu Q)}
(b^\dagger(\v{k})b(\v{k})e^{ik(x-x')}+a^\dagger(\v{k})a(\v{k})e^{-ik(x-x')})|\{n\}\ra\nonumber
\eea
which in turn gives
\bea
iD^{-+}(x,x')&=&\frac{\Tr[e^{-\beta(H-\mu Q)}\phi(x)\phi^\dagger(x')]}{\Tr[e^{-\beta(H-\mu Q)}]}\nn
&=&iD^{-+}_{T=0}(x,x')+\int_{\tilde{\v{k}}}e^{ik(x-x')}\frac{\sum_nne^{-\beta(\epsilon_\v{k}+\mu)}}{\sum_ne^{-\beta(\epsilon_\v{k}+\mu)}}
+\int_{\tilde{\v{k}}}e^{-ik(x-x')}\frac{\sum_nne^{-\beta(\epsilon_\v{k}-\mu)}}{\sum_ne^{-\beta(\epsilon_\v{k}\mu)}}\nn
&=&iD^{-+}_{T=0}(x,x')+\int_ke^{-ik(x-x')}\left(\frac{2\pi\delta(k^2-m^2)\Theta(-k^0)}{e^{\beta(\epsilon_\v{k}+\mu)}-1}
+\frac{2\pi\delta(k^2-m^2)\Theta(k^0)}{e^{\beta(\epsilon_\v{k}-\mu)}-1}\right).
\eea
The modification due to the environment is the same in $\Tr[e^{-\beta(H-\mu Q)}\phi^\dagger(x')\phi(x)]$
as in Eq. \eq{mpcb}, leading to the same conclusion as in the case of a neutral boson,
namely that the only modification of the environment is $D^{\pm\mp}=D^{\pm\mp}_{T=0}+\Delta D$, and
\be\label{cbdpm}
i\Delta D_{ik}=-i\frac{2\pi\delta(k^2-m^2)\Theta(-k^0)}{e^{\beta(\epsilon_\v{k}+\mu)}-1}
-i\frac{2\pi\delta(k^2-m^2)\Theta(k^0)}{e^{\beta(\epsilon_\v{k}-\mu)}-1}.
\ee

\subsection{Fermions}
The field operator 
\be
\psi(x)=\frac1V\sum_{\v{k}\alpha}\frac{m}{\omega_k}[b_\alpha(\v{k})u^{(\alpha)}(\v{k})e^{-ikx}
+d^\dagger_\alpha(\v{k})v^{(\alpha)}(\v{k})e^{ikx}]
\ee
and the anti-commutation relations
\be
\{b_\alpha(\v{p}),b^\dagger_\beta(\v{p}')\}=\{d_\alpha(\v{p}),d^\dagger_\beta(\v{p}')\}
=V\frac{\omega_p}{m}\delta_{\alpha,\beta}\delta_{\v{p},\v{p}'}
\ee
give
\bea
\Tr[e^{-\beta(H-\mu Q)}\psi(x)\psib(x')]&=&\sum_{\{n\}}\sum_{\v{k},\v{k}'}\frac{m^2}{V^2\omega_\v{k}\omega_{\v{k}'}}\la\{n\}|e^{-\beta(H-\mu Q)}\nn
&&(d^\dagger(\v{k})d(\v{k}')v^{(\alpha)}(\v{k})\bar v^{(\alpha)}(\v{k}')e^{ikx-ik'x'}
+b(\v{k})b^\dagger(\v{k}')u^{(\alpha)}(\v{k})\bar u^{(\alpha)}(\v{k}')e^{ik'x'-ikx})|\{n\}\ra\nn
&=&\sum_{\{n\}}\sum_\v{k}\frac{m}{2V^2\omega^2_\v{k}}\nn
&&\la\{n\}|e^{-\beta(H-\mu Q)}(d^\dagger(\v{k})d(\v{k})(k\br-m)e^{ik(x-x')}+b(\v{k})b^\dagger(\v{k})(k\br+m)e^{-ik(x-x')})|\{n\}\ra\nn
&=&\sum_{\{n\}}\sum_\v{k}\frac{1}{2V\omega_\v{k}}(k\br+m)e^{-ik(x-x')}\la\{n\}|e^{-\beta(H-\mu Q)}|\{n\}\ra_{E>0}\nn
&&+\sum_{\{n\}}\sum_\v{k}\frac{m}{2V^2\omega^2_\v{k}}\nn
&&\la\{n\}|e^{-\beta(H-\mu Q)}(d^\dagger(\v{k})d(\v{k})(k\br-m)e^{ik(x-x')}-b^\dagger(\v{k})b(\v{k})(k\br+m)e^{-ik(x-x')})|\{n\}\ra\nn
&=&\int_{\tilde{\v{k}}}(k\br+m)e^{-ik(x-x')}\prod_\v{k}(1+e^{-\beta(\epsilon_\v{k}-\mu)})\nn
&&+\sum_\v{k}\frac{1}{2V\omega_\v{k}}(e^{-\beta(\epsilon_\v{k}+\mu)}(k\br-m)e^{ik(x-x')}
-e^{-\beta(\epsilon_\v{k}-\mu)}(k\br+m)e^{-ik(x-x')}).
\eea
This result, together with the analogous one for $\Tr[e^{-\beta(H-\mu Q)}\psib(x')\psi(x)]$ yield
\be
G^{\pm\mp}=-(i\partial\br+m)(D^{\pm\mp}_{T=0}-\Delta D),
\ee
where the last term is given by Eq. \eq{cbdpm}.

\subsection{Summary}
The different propagators encountered in the discussion presented above can be summarized as
\be\label{bosonctp}
D_k=\begin{pmatrix}\frac{1}{k^2-m^2+i\epsilon}&-2\pi i\delta(k^2-m^2)\Theta(-k^0)\cr
-2\pi i\delta(k^2-m^2)\Theta(k^0)&-\frac{1}{k^2-m^2-i\epsilon}\end{pmatrix}
-i2\pi\delta(k^2-m^2)[\Theta(k^0)n^+_b+\Theta(-k^0)n^-_b]\begin{pmatrix}1&1\cr1&1\end{pmatrix}
\ee
and
\be\label{fermionctp}
G_k=(k\br+m)\biggl[\begin{pmatrix}\frac{1}{k^2-m^2+i\epsilon}&2\pi i\delta(k^2-m^2)\Theta(-k^0)\cr
2\pi i\delta(k^2-m^2)\Theta(k^0)&-\frac{1}{k^2-m^2-i\epsilon}\end{pmatrix}
+2\pi i\delta(k^2-m^2)[\Theta(k^0)n^+_f+\Theta(-k^0)n^-_f]\begin{pmatrix}1&-1\cr-1&1\end{pmatrix}\biggr]
\ee
where the occupation numbers
\bea\label{distrfncs}
n^\pm_b&=&\frac1{e^{\beta(\epsilon_\v{k}\mp\mu)}-1}\nn
n^\pm_f&=&\frac1{e^{\beta(\epsilon_\v{k}\mp\mu)}+1}
\eea
are used.

\end{document}